\documentclass[preprint,showkeys,aps,doublespace]{revtex4}
\usepackage[dvips]{graphicx}
\usepackage{setspace}
\usepackage{amsmath}
\usepackage{amsfonts}
\usepackage{amssymb,color}
\usepackage{subfigure}
\usepackage{graphicx,color}
\usepackage{ifpdf}
\usepackage[latin1]{inputenc}

\begin{document}

\title{Driven Widom-Rowlinson lattice gas}
\author{ Ronald Dickman$^1$\footnote{email: dickman@fisica.ufmg.br}
and R.K.P. Zia$^2$\footnote{email: rkpzia@vt.edu} }

\address{
$^1$ Departamento de F\'{\i}sica and
National Institute of Science and Technology for Complex Systems,
ICEx, Universidade Federal de Minas Gerais,
C. P. 702, 30123-970 Belo Horizonte, Minas Gerais - Brazil\\
$^2$ Center for Soft Matter and Biological Physics, Department of Physics,
Virginia Polytechnic Institute \& State University, Blacksburg, VA 24061, USA
}

\date{\today }

\begin{abstract}
In the Widom-Rowlinson lattice gas, two particle species (A, B)
diffuse freely via particle-hole exchange, subject to both on-site
exclusion and prohibition of A-B nearest-neighbor pairs. As an athermal
system, the overall densities are the only control parameters. As the
densities increase, an entropically driven phase transition occurs, leading to
ordered states with A- and B-rich domains separated by hole-rich interfaces.
Using Monte Carlo simulations, we analyze the effect of imposing a drive on
this system, biasing particle moves along one direction. Our study parallels
that for a driven Ising lattice gas -- the Katz-Lebowitz-Spohn (KLS) model, which
displays atypical collective behavior, e.g., structure factors with
discontinuity singularities and ordered states with domains only parallel to
the drive. Here, other novel features emerge, including structure factors
with \textit{kink} singularities (best fitted to $\left\vert q\right\vert $),
maxima at non-vanishing wavevector values, oscillating correlation
functions, and ordering into \textit{multiple} striped domains
\textit{perpendicular} to the drive, with a preferred wavelength depending on density and drive intensity.
Moreover, the (hole-rich) interfaces between the domains are statistically rough
(whether driven or not), in sharp contrast with those in the KLS model, in which the drive
suppresses interfacial roughness. Defining a novel order parameter (to account for the
emergence of multistripe states), we map out the phase diagram in the density-drive plane
and present preliminary evidence for a critical phase in this driven lattice gas.

\end{abstract}

\maketitle

\address{
$^1$Departamento de F\'{\i}sica and
National Institute of Science and Technology for Complex Systems,\\
ICEx, Universidade Federal de Minas Gerais, \\
C. P. 702, 30123-970 Belo Horizonte, Minas Gerais - Brazil\\
$^2$Center for Soft Matter and Biological Physics, Department of Physics,\\
Virginia Polytechnic Institute \& State University,\\
Blacksburg, VA 24061, USA\\
Department of Physics and Astronomy, Iowa State University, Ames, IA, 50011, USA
}

\section{Introduction}

Driven lattice gases have played a central role in far-from-equilibrium
thermodynamics and statistical mechanics, in particular in the study of
phase transitions \cite{schmittmann-zia,marro} and steady-state
thermodynamics (SST) \cite{ST}. In this context, most studies have focussed on the lattice gas with attractive
nearest-neighbor (NN) interactions pioneered by Katz, Lebowitz, and Spohn \cite{KLS} (KLS).
In equilibrium, this model is the lattice gas version of the Ising model \cite{LeeYang}.
Under a drive favoring particle motion along one axis (and suppressing
motion in the opposite sense), the KLS model
exhibits a variety of remarkable collective behaviors \cite%
{schmittmann-zia,marro}, such as generic long-range correlations
\cite{LRC,Onuki79,Onuki81,Kirkpatrick16},
`negative response' \cite{ZiaPM}, non-Ising critical properties \cite{JSLC},
anisotropic phase separation \cite{KLS}, and anomalous interfacial
fluctuations \cite{leung88,InterfaceCorr}. More recently, it has been
extensively studied in connection with the difficulties in formulating a
zeroth law of thermodynamics for systems driven into steady states far from
equilibrium \cite{pradhan,sst1,sst3}. Another model of interest is the
driven lattice gas with NN exclusion (NNE lattice gas) \cite{drnne}, which
exhibits a complex pattern of phase ordering and jamming, and has also
proved useful in efforts to define a nonequilibrium chemical potential \cite%
{sst2}. The NNE lattice gas is an athermal model exhibiting a phase
transition to sublattice ordering. While athermal models are convenient in
that they involve only a single effective parameter (the dimensionless
chemical potential $\mu ^{\ast }=\mu /k_{B}T$), the NNE lattice gas does not
exhibit macroscopically distinguishable phases. An athermal model that does
exhibit phase separation is the lattice Widom-Rowlinson (WR) model or
Widom-Rowlinson lattice gas (WRLG) \cite{wrlg}. It is likely the simplest
model with purely repulsive interactions to undergo phase separation. This
study is devoted to a driven WRLG.

The equilibrium WRLG was introduced in \cite{wrlg}, as a discrete-space
version of the original (continuous-space) Widom-Rowlinson model \cite%
{WR70,RW82}. Each site may be vacant, or occupied by a particle belonging to
species A or species B. Nearest-neighbor A-B pairs are prohibited, as is
multiple occupancy of a site. (The latter restriction departs from the
original, continuous-space WR model, in which particles belonging to the
same species do not interact at all.) In \cite{wrlg} it is shown that the
WRLG (with equal densities of A and B particles, $\rho _{A}=\rho _{B}\equiv
\rho /2$) exhibits a continuous phase transition between a disordered,
low-density phase and coexisting A- and B-rich phases at a critical density $%
\rho _{c}$. Best estimates for the critical density are 0.618(1) and
0.3542(1) on the square and
simple cubic lattices, respectively.
(Figures in parentheses denote statistical uncertainties.)
Results on scaling behavior \cite{wrlg}
support an Ising-like critical point, as expected on grounds of symmetry.
Subsequent studies of the WRLG have focussed on multi-species versions \cite%
{Nielaba-Lebowitz}. Here we study the \textit{driven} WRLG: as in the KLS
model, and the driven NNE lattice gas, particle motion is favored along one
direction, and suppressed in the opposite sense, as if an electric or other
external field were acting on the particles. To model a nonequilibrium
steady state, it is essential that the system have periodic boundaries along
the drive direction. (For impenetrable boundaries the system corresponds to
an \textit{equilibrium} fluid subject to a simple linear gravitational
potential.)

In the NNE lattice gas above the critical density,
the sublattice-ordered phases are equivalent under a unit translation in
the $x$ or $y$ direction, whereas the symmetry connecting the A- and B-rich phases
in the WRLG is global exchange of A and B particles. In the
driven NNE lattice gas with first- and second-neighbor hopping (the model
studied by Szolnoki and Szabo, \cite{szolnoki-szabo02}), particles are free to hop between
sublattices, so there is global sublattice ordering rather than spatial
separation into two phases; in this case no stripe pattern is observed.

In the single-component lattice gas with finite \textit{repulsive} NN
interactions, (equivalent in equilibrium to the antiferromagnetic Ising
model), the drive destroys sublattice ordering \cite{szabo94,szabo96}. For
infinite NN repulsion, i.e., the nearest-neighbor exclusion (NNE) model,
phase separation under a drive is possible, as well as formation of
dynamically arrested regions, if the dynamics allows NN hopping only \cite%
{drnne}. Including hopping to second neighbors changes the phase behavior:
instead of separation into high- and low-density regions, there is a
continuous, Isinglike transition to a phase with sublattice ordering, as in
equilibrium \cite{szolnoki-szabo02}. In the present work, the
dynamics includes both NN and second-neighbor hopping.

There are two principal motivations for studying the driven WRLG. First,
aside from the examples already cited, not many driven systems exhibiting
phase separation have been examined in detail; the lack of attractive
interactions leads to a very different sort of phase ordering under a drive,
as compared with the KLS model.
In this context, we should mention that the WRLG is a spin-1 or 3-state system,
a class that includes well known systems such as the Potts \cite{Potts}
and Blume-Emery-Griffiths \cite{BEG} models.
Some \textit{driven} 3-state models have been investigated previously \cite{Threestate},
but none display the novel properties of the system we present here.
Second, it is valuable to have further
examples of model systems for testing SST.

There are two main foci in the present work: anomalous correlations (even in
the disordered phase) and remarkably rich varieties of phase segregated
order. Our principal source of information is Monte Carlo simulation of the
model on the square lattice, which we complement with simple mean-field
theory approaches. The first issue is best displayed in terms of the
structure factor $S(q_{||},q_{\perp })$, with the subscripts denoting
wavevector components along, and perpendicular to, the drive. In addition to
displaying a discontinuity singularity at the origin [$S(q_{||}\rightarrow
0,0)\neq S(0,q_{\perp }\rightarrow 0)$] as observed in similar driven
diffusive systems \cite{Huang93,Zia95,Korniss97,Praestgaard00,Vidigal07,Greulich08},
$S(q_{||}\rightarrow 0,0)$ exhibits a \textit{kink} singularity,
best fitted to $\left\vert q_{||}\right\vert $! This
novel feature is accompanied by a \textit{maximum} of $S(q_{||},0)$ located
at some non-vanishing $q_{||}$ value, associated with periodic
correlations of composition fluctuations along the drive direction. Let us emphasize that these properties
are present even at low densities (for which the composition is spatially
uniform), in stark contrast to the familiar, analytic, Ornstein-Zernike form
$S(q_{||},0)\propto 1/(\tau +q_{||}^{2})$. Under the drive, composition
variations are characterized by a preferred wavelength $\lambda (\rho ,p)$,
where $p$ denotes the intensity of the drive.
The second focus is a phase diagram in the density-drive plane, as we report
the emergence of \textit{multiple} stripes of A- and B-rich phases [of
wavelength $\lambda (\rho ,p)$], oriented \textit{perpendicular} to the
drive! By contrast, in the KLS model, the particle-rich (or hole-rich)
regions form a \textit{single} stripe \textit{parallel} to the drive
(especially in systems with $O\left( 1\right) $ aspect ratios). As in the
KLS model, long-lived metastable states are observed as control parameters
are changed across phase boundaries, a phenomenon reminiscent of hysteresis.

The remainder of this paper is organized as follows. In the next section we
define the model.
In Sec. III we examine the low-density, disordered phase, and discuss mean-field
and field-theoretic approaches.  Sec. IV reports our results for local and global
ordering, correlations in the ordered phase, and interface roughness,
followed by a summary and discussion of our findings in Sec. V .

\section{Model}

The WRLG is defined on a lattice of $L^{d}$ sites, each of which may be
empty or occupied by a particle of type A or of type B; multiple occupancy
is forbidden. It is convenient to introduce occupation variables $\sigma
_{i}=0$, $1$, or $-1$ for site $i$ vacant, occupied by an A particle, or
occupied by a B particle, respectively. To model the NN repulsion, if $%
\sigma _{i}=\pm 1$, none of its nearest neighbors may have $\sigma =\mp 1$.
In equilibrium, the model may be studied in the canonical ensemble, i.e.,
with fixed numbers of A and B particles, $N_{A}$ and $N_{B}$, respectively,
or in the grand canonical ensemble, in which the associated chemical
potentials, $\mu _{A}^{\ast }$ and $\mu _{B}^{\ast }$ are fixed. Here we
limit our attention to the case of square lattice ($d=2$). Thus, a site is
labeled by integers $\left( n_{x},n_{y}\right) \in $ $\left( \left[ 1,L_{x}%
\right] ,\left[ 1,L_{y}\right] \right)$,
while $\sigma _{i}$ is denoted by $\sigma
\left( \vec{x}\right) $. Unless explicitly noted otherwise, our study
focuses on systems with $L_{x}=L_{y}=L$. For simplicity, we focus on systems
with fixed $N_{A}=N_{B}=N$, and choose the main control parameter to be the
overall density: $\rho =2N/\left( L_{x}L_{y}\right) $.

It is well to recall that the equilibrium thermodynamics of athermal systems
is determined exclusively by maximization of entropy, subject to whatever
constraints apply. Thus phase separation in the WR model occurs at densities
high enough that the configurational entropy of the phase-separated,
inhomogeneous system is higher than that of a homogenous one. As in the
freezing of the hard-sphere fluid, a higher entropy is associated with a
nominally more ordered phase because the disordered phase possesses
relatively few configurations, due to excluded-volume constraints. To what
extent such considerations apply to \textit{driven} athermal systems is an
open question: under a drive, the probability distribution is no longer
uniform on configuration space, and the very definition of a thermodynamic
entropy is in general unclear.

The \textit{driven} WRLG is a stochastic process with a Markovian dynamics
defined via a set of transition rates $w(\mathcal{C}^{\prime },\mathcal{C})$%
, from configuration $\mathcal{C}$ (i.e., the set $\left\{ \sigma
_{i}\right\} $) to configuration $\mathcal{C}^{\prime }$. Transitions occur
via single-particle hopping, so that for $w(\mathcal{C}^{\prime },\mathcal{C}%
)$ to be nonzero, configurations $\mathcal{C}$ and $\mathcal{C}^{\prime }$
must differ by the exchange of a particle-hole pair. Specifically, each
particle attempts to hop to a first or second neighbor, that is, a
displacement of $\left(\Delta n_{x}, \Delta n_{y} \right) $ with $\Delta n_{x}$,
$\Delta n_{y} \in \{-1,0,1\}$ (excluding the no-hop case $\Delta n_{x} = \Delta n_{y} =0$). Any hopping move
that results in a configuration satisfying the prohibition against NN A-B
pairs is accepted. The attempt rates are parameterized in terms of $p\in
\lbrack -1,1]$ (propensity to move along $+x$) and $a\in \lbrack 0,1]$
(tendency to leave $x$ unchanged):

\begin{equation}
w(\Delta n_{x}, \Delta n_{y} )=\left\{
\begin{array}{ll}
\frac{(1\!-\!a)(1\!\pm \!p)}{6},\;\; \Delta n_{x} =\pm 1,\text{ any } \Delta n_{y} &  \\
\frac{a}{2},\;\;\;\;\;\;\;\;\;\;\;\; \Delta n_{x} =0, \Delta n_{y} =\pm 1 &  \\
0,\;\;\;\;\;\;\;\;\;\;\;\;\Delta n_{x} =0= \Delta n_{y} &
\end{array}%
\right. \,.  \label{hoprates}
\end{equation}

For $p=0$ and $a<1$, detailed balance is satisfied, and the stationary
distribution of the Markov process corresponds to the equilibrium
distribution, i.e., all allowed configurations equally likely. The hopping
rates are isotropic for $p=0$ and $a=1/4$. (The special case of $a=1$, in
which particles are restricted a given column, is not considered here.) As $%
p $ is varied from zero to unity, the process interpolates from a case with
symmetric hopping in $x$ (equilibrium) to one with maximal asymmetry, with
no hopping in the $-x$ direction.
We consider a dynamics with next-nearest-neighbor as well as nearest-neighbor
hopping to avoid the possibility of nongeneric patterns or dynamic arrest associated
with a restriction to NN hopping \cite{szolnoki-szabo02}, and because an extended set of
hopping moves favors ergodicity.  Our choice of equal hopping rates for jumps with
the same value of $\Delta n_x$ is motivated by simplicity.
In the simulation studies reported here,
we set $a=0$ and only comment on some runs with $a=1/4$.

\subsection{Simulations}

We perform Monte Carlo (MC) simulations the driven WRLG on the square
lattice, using rectangular systems of $L_{x}\times L_{y}$ sites with
periodic boundaries in both directions. Unless otherwise noted, we use $%
L_{x}=L_{y}=L$, with $L$ ranging from 30 to 400. We count one MC step as $N$
attempted particle moves. Studies of stationary properties use $2-14\times
10^{7}$ MC steps, preceded by an initial period of $10^{7}$ MC steps, which
we found to be sufficient for relaxation. Averages and uncertainties
(standard deviation of the mean) are calculated over 10-30 independent
realizations.

\section{Anomalies in the homogeneous phase: Singular structure factors and
correlations}

If the density $\rho $ is sufficiently low, the WRLG remains homogeneous whether driven
or not. Simulations of the equilibrium case show this disordered phase to
prevail for $\rho <$ $\rho _{c}\simeq 0.618\left( 1\right) $. As the drive
parameter $p$ is increased from zero to unity, the change in $\rho _{c}$ is
rather modest, attaining a value of about $0.76(1)$ for $p=1$. By contrast,
the critical temperature of the driven KLS model is found to increase by as
much as $40\%$ (in $d=2$) from equilibrium \cite{KLS}. In this section, we
focus our attention on the low-density disordered state, in which we find
anomalies in the correlation functions and structure factors, beyond those
observed in the KLS system.

\subsection{Simulation studies of the structure factor and correlation
functions}

One standard way to characterize collective behavior is through correlation
functions ($G$) and their Fourier transforms, the structure factors ($S$).
For simplicity, we consider only equal-time, two-point correlations;
since there are two species of particles, these are $2\times 2$ matrices.
Instead of the densities of the two particle species, $\rho _{A,B}\left(
\vec{x}\right) $, symmetry guides us to consider their sum and difference,
which will be referred to as \textquotedblleft mass\textquotedblright\ and
\textquotedblleft charge\textquotedblright\ densities
\begin{eqnarray*}
\rho \left( \vec{x}\right) &\equiv &\rho _{A}\left( \vec{x}\right) +\rho
_{B}\left( \vec{x}\right), \\
\psi \left( \vec{x}\right) &\equiv &\rho _{A}\left( \vec{x}\right) -\rho
_{B}\left( \vec{x}\right),
\end{eqnarray*}%
respectively \cite{notation}.
While this notation is
unnecessary for reporting simulation data (as they are just, respectively, $%
\left\vert \sigma \left( \vec{x}\right) \right\vert $ and $\sigma \left(
\vec{x}\right) $ at each $\vec{x}$), it sets the stage for a field-theoretic
approach, the simplest of which will be discussed below. The restrictions of
our simulations correspond to the constraints $\langle \left\vert \sigma
\left( \vec{x}\right) \right\vert \rangle =\rho $ and $\langle \sigma \left(
\vec{x}\right) \rangle =0$.

In terms of these densities, the $2\times 2$ matrix of correlations consists
of $G_{\rho \rho }$ and $G_{\psi \psi }$. The cross correlation, $G_{\rho
\psi }=G_{\psi \rho }$, is the difference $G_{AA}-G_{BB}$ which vanishes
(statistically) in the symmetric systems we study. For simply, we write $%
G_{\rho \rho ,\psi \psi }$ as $G_{\rho }$ and $G_{\psi }$. In the steady
state, translational invariance dictates that these are functions of $\vec{r}
$, the distance between the two points in question. Thus,

\begin{eqnarray}
G_{\rho }(\vec{r}) &\equiv &\langle \left\vert \sigma \left( \vec{x}\right)
\right\vert \left\vert \sigma \left( \vec{x}+\vec{r}\right) \right\vert
\rangle -\rho ^{2}  \notag \\
G_{\psi }(\vec{r}) &\equiv &\langle \sigma \left( \vec{x}\right) \sigma
\left( \vec{x}+\vec{r}\right) \rangle  \notag
\end{eqnarray}%
are independent of $\vec{x}$; to improve statistics, we average over $\vec{x}
$. When we form the covariance or connected correlation function
by subtracting the associated expectations, $\left\langle \sigma \right\rangle
\left\langle \sigma \right\rangle $, these $G$'s assume different values at
the origin: $G_{\rho }(\vec{0})=\rho \left( 1-\rho \right) $ \textit{vs.} $%
G_{\psi }(\vec{0})=\rho $, since $\sigma ^{2}\left( \vec{x}\right)
=\left\vert \sigma \left( \vec{x}\right) \right\vert ^{2}=\left\vert \sigma
\left( \vec{x}\right) \right\vert $. Meanwhile, the sum $\sum_{\vec{r}}G(%
\vec{r})$ vanishes for both functions. The structure factors are the Fourier
transforms%
\begin{equation}
S\left( \vec{q}\right) =\sum_{\vec{r}}G(\vec{r})e^{i\vec{q}\cdot \vec{r}}
\label{SFdef}
\end{equation}%
By averaging over $\vec{x}$ to obtain $G$, we also have, e.g., $S_{\psi
}=L^{-2}\sum_{\vec{x},\vec{r}}\langle \sigma \left( \vec{x}\right) \sigma
\left( \vec{x}+\vec{r}\right) \rangle e^{i\vec{q}\cdot \vec{r}}=\sum_{\vec{x}%
,\vec{x}^{\prime }}\langle \sigma \left( \vec{x}\right) e^{-i\vec{q}\cdot
\vec{x}}\sigma \left( \vec{x}^{\prime }\right) e^{i\vec{q}\cdot \vec{x}%
^{\prime }}\rangle $, which is $L^{-2}$ times $\left\langle \left\vert \sum_{%
\vec{x}}\sigma \left( \vec{x}\right) e^{i\vec{q}\cdot \vec{x}}\right\vert
^{2}\right\rangle $, the average over a run of the power spectrum \cite{beam},
of each configuration $\sigma \left( \vec{x}\right) $. Finally, the constraints
on $G$ translate into $\sum_{\vec{q}}S_{\rho }\left( \vec{q}\right) =\rho
\left( 1-\rho \right) L^{2}$, $\sum_{\vec{q}}S_{\psi }\left( \vec{q}\right)
=\rho L^{2}$, and $S_{\rho ,\psi }\left( \vec{0}\right) =0$. These serve as
useful checks on simulation data. In addition, the last constraint implies
that, unlike the ordinary Ising model, $S\left( \vec{0}\right) $ cannot
serve as a susceptibility-like quantity for detecting a second-order
critical point. Instead, our $S$ assumes its maximum value at some non-vanishing $\vec{q}$.
The maximum value behaves like the susceptibility of ferromagnetic systems, i.e.,
remaining finite (as $L \to \infty$) in the disordered phase.
For large enough $\rho$, as we report in Sec.~IV,
this value diverges as $L^2$ and so is an ideal candidate for an order parameter.
If the transition turns out to be critical (i.e., behaves much like second-order
transitions in equilibrium statistical systems), then the large-$L$ behavior of
$S(q_{max})$ can provide us with a critical exponent.

In general, the two-point \textit{equal-time} correlation, $G\left( \vec{r}%
\right) $, must be symmetric under parity ($\vec{r}\Leftrightarrow -\vec{r}$%
) and statistically symmetric under the reflections ($%
x\Leftrightarrow -x$ and $y\Leftrightarrow -y$). In equilibrium, there is an
additional symmetry, the $x\Leftrightarrow y$ exchange (especially for
samples with $L_{x}=L_{y}$). By contrast, the drive should break the last
symmetry. Thus, in addition to studying $G$ and $S$ in the whole plane, we
will pay special attention to the specific directions along and
perpendicular to the drive.  In particular, we will
consider, e.g.,

\begin{equation}
G_{\psi ||}(r)\equiv \langle \sigma \left( \vec{x}\right) \sigma \left( \vec{%
x}+r\hat{x}\right) \rangle ;~~G_{\psi \perp }(r)\equiv \langle \sigma \left(
\vec{x}\right) \sigma \left( \vec{x}+r\hat{y}\right) \rangle
\label{G-on-axes}
\end{equation}%
and

\begin{equation}
S_{\psi ||}(k)\equiv S_{\psi }\left( q_{||}=k,0\right) ;~~S_{\psi \perp
}(k)\equiv S_{\psi }\left( 0,q_{\perp }=k\right)  \label{S-on-axes}
\end{equation}%
Let us call attention to the different roles played by the quantities $\vec{q}$
and $k$. The former is a wave vector, being the conjugate
to the vector $\vec{x}$, with components $\left(  q_{||},q_{\perp}\right)  $.
The latter is a wavenumber, measuring the magnitude of $\vec{q}$ along
\textit{one} of two axes, in the same spirit that $r$ is the magnitude of
$\vec{x}$ along \textit{one} of the axes. Thus, $S_{\psi||}(k)$ and
$G_{\psi||}(r)$ are \textit{not} simply related to each other
by Fourier transforms. Instead, $S_{\psi ||}\left( k\right) $ is the
transform associated with the \textit{average} charge density over a column $%
\bar{\psi}_{||}\left( x\right) \equiv L^{-1}\sum_{y}\sigma \left( x,y\right)
$ and its correlation function ${C_{\psi ||}(r)=\langle \bar{\psi}%
_{||}\left( x\right) \bar{\psi}_{||}\left( x+r\right) \rangle }$. Similarly,
we can define the transverse counterparts: $\bar{\psi}_{\bot }\left(
y\right) \equiv L^{-1}\sum_{x}\sigma \left( x,y\right) $ and $C_{\psi \bot
}\left( r\right) {=}{\langle \bar{\psi}_{||}\left( y\right) \bar{\psi}%
_{||}\left( y+r\right) \rangle }$. (Of course, translational invariance of
the steady state provides the $x$ or $y$ independence of the $C$'s.) We will
not discuss these functions explicitly below, but focus on the $S$'s, as the
latter contain the same information about the system, displayed in a better
form.

With this setup, let us turn to the remarkable behavior revealed by
simulations. For simplicity, we focus mainly on $\psi $ and occasionally
mention results for $\rho $. Thus, we drop the subscripts on $G_{\psi }$ and
$S_{\psi }$ as well. Considering first the $S$'s, we report results found
mainly in a $a=0,p=1,\rho =0.4,L=100$ system. The reason for this choice is
that the novel and most interesting features are best displayed here.

In Fig.~1, we show $S\left( \vec{q}\right) $ and immediately
notice a feature quite distinct from the structure factor of the KLS model,
namely, the presence of \textquotedblleft twin peaks.\textquotedblright\
Despite being in the disordered, homogeneous phase, such prominent
structures represent a major deviation from the simple Lorentzian $1/(\tau +%
\vec{q}^{2})$ form of the equilibrium WRLG, as well as the discontinuity
singularity at ${\vec q}=0$ found in the KLS model \cite{KLS,schmittmann-zia},
$\left( \nu _{x}q_{x}^{2}+\nu
_{y}q_{y}^{2}\right) /\left( \tau _{x}q_{x}^{2}+\tau _{y}q_{y}^{2}+O\left(
q^{4}\right) \right) $. While the data show $S_{\bot }\left( k\right) $
decreasing monotonically as $k$ increases from zero, $S_{||}\left(
k\right) $ rises to a maximum at wavenumber $9$ before dropping. Such a peak
is indicative of a preferred wavelength induced by the drive: $\lambda
\equiv 2\pi /k_{\max }$.

\begin{figure}[h]
\label{Sp1r4L100} \center
\subfigure[]{\includegraphics[scale=1.05]{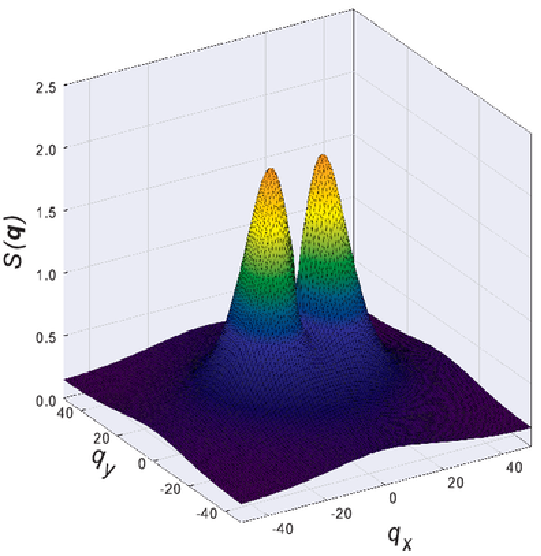}} \qquad %
\subfigure[]{\includegraphics[scale=0.55]{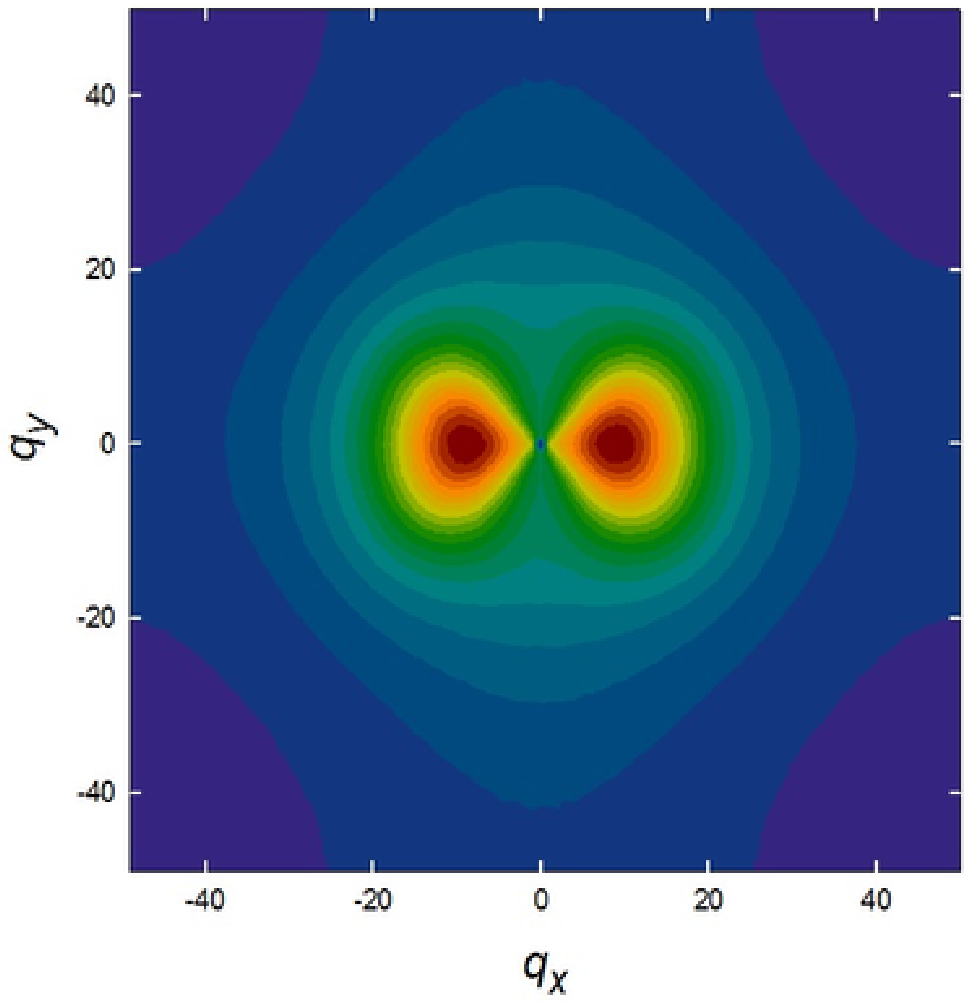}}
\caption{(Color online) Structure factor for $\rho=0.4$, $p=1$, $a=0$, system size $L=100$.
Left: three-dimensional surface plot; right: contour plot.}
\end{figure}

\begin{figure}[h!]
\centering
\includegraphics[scale=0.7]{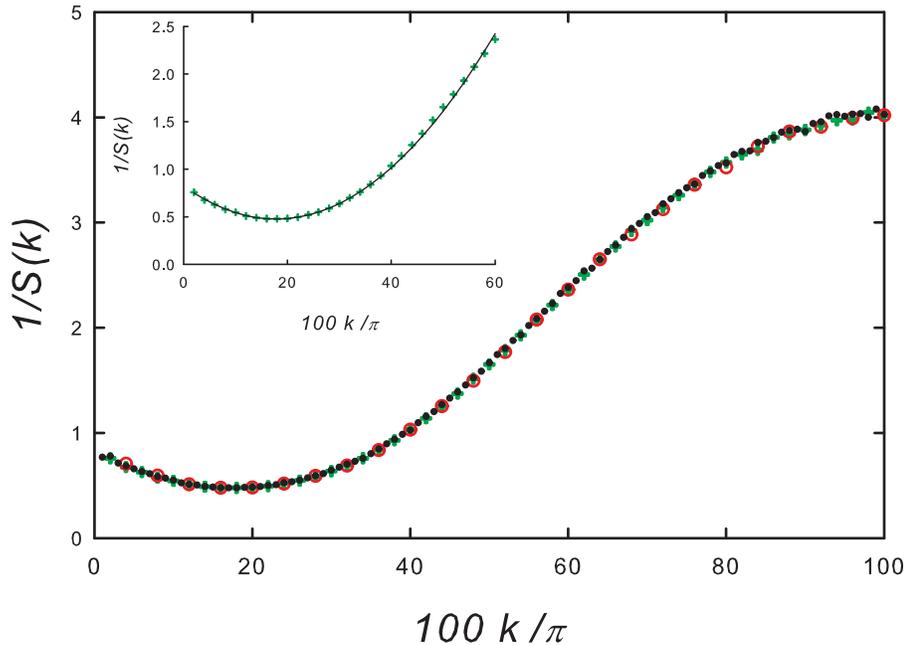}
\caption{(Color online) Reciprocal of structure factor for system sizes $L=50$ (circles), 100 ($+$ signs)
and 200 (dots); other
parameters as in Fig.~1. Inset: data for $L=100$ (points) compared with quadratic fit (curve) to 17 points nearest
the minimum.}
\label{FFS}
\end{figure}

\begin{figure}[h!]
\centering
\includegraphics[scale=0.7]{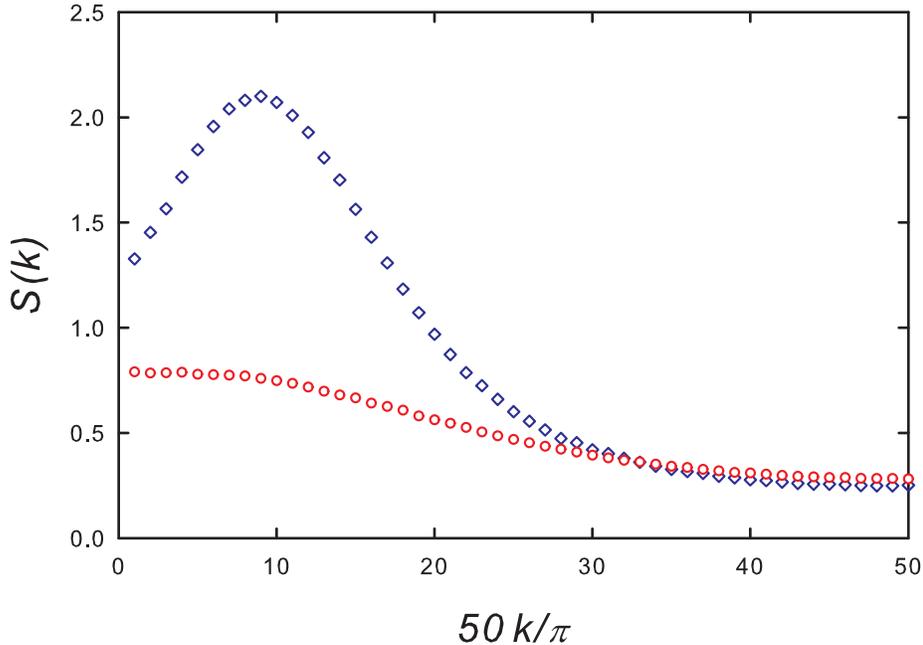}
\caption{(Color online) Structure factors $S(k_x,0)$ (diamonds) and $S(0,k_y)$ (circles) for
parameters as in Fig.~1.}
\label{sparsprp}
\end{figure}

To be more confident of this unusual behavior in $S_{||}\left( k\right) $,
we considered finite size effects, illustrated in Fig.~\ref{FFS}. As the
figure shows, the data points for $L=50$ and $200$ fall well within
statistical errors of the $L=100$ sample, so that, for this
density and drive, the thermodynamic limit is well represented by the $L=100$
system. Thus, we see that the small-$k$ region of $const./S_{||}$ is well fit by
$\tau +\left( \left\vert k\right\vert -k_{\max }\right) ^{2}$.
Alternatively, we can write $S_{||}\left( k\right) \propto 1/\left[ 1+\left(
\left\vert k\right\vert -k_{\max }\right) ^{2}\xi _{||}^{2}+O\left(
k^{4}\right) \right] $, which displays the presence of another length, $\xi
_{||}$, a candidate for a \textquotedblleft correlation
length.\textquotedblright\ Note that $S_{||}\left( k\right) $ must be even
in $k$, as the equal-time two-point correlations in the steady state must be
even in $x$. Thus, the presence of peaks symmetric about $k=0$ is
inevitable. Although, in principle, $\left( k^{2}-k_{\max }^{2}\right) ^{2}$
satisfies these symmetry conditions, Fig.~\ref{sparsprp} (diamonds) shows
that there is, without doubt, a \textit{kink singularity} ($\propto \left\vert
k\right\vert $) at $k=0$, so that the quadratic expression proposed above is a
much better fit to $1/S_{||}$. In this regard, though a kink
singularity is not necessary to generate period structures, the observed
oscillations are best modeled by one. The inset of Fig.~\ref{FFS} shows that the
data for $1/S_{||}$ are well fit by a quadratic expression,
$0.475+1.11\left( \left\vert
k\right\vert -k_{\max }\right) ^{2}$, with
\begin{equation*}
k_{\max }\cong 0.56;~~\lambda \cong 11.
\end{equation*}%

\noindent This form implies that $\xi _{||}\simeq \sqrt{1.11/0.475}\simeq 1.5$.
Meanwhile, Fig. \ref{sparsprp} (circles) shows that $S_{\bot }$ is
consistent with the ordinary Lorenztian form. Of course, the drive induces
anisotropy, so that $\xi _{\bot }\neq \xi _{||}$ in general. As in the KLS model, only
one of the correlation lengths, $\xi_{||}$, diverges as the critical point is approached,
as discussed below.
Our best phenomenological estimate for the structure factor
(for small $\vec{k}$, for this sample) is%
\begin{equation}
S\left( \vec{q}\right) =\frac{\nu _{||}q_{||}^{2}+\nu _{\bot }q_{\bot }^{2}}{%
\tau _{||}q_{||}^{2}+\tau _{\bot }q_{\bot }^{2}+q_{||}^{2}\left( \left\vert
q_{||}\right\vert -k_{\max }\right) ^{2}+2\gamma _{\times }q_{||}^{2}q_{\bot
}^{2}+\gamma _{\bot }q_{\bot }^{4}+...}  \label{PhenoS}
\end{equation}%
While such a form may seem unusual, it is based on a stochastic field theory
which has proven successful in describing the KLS model \cite{JSLC} (the numerator and
denominator being associated with the noise and deterministic part of a
Langevin equation for the density field). The main novel feature here is the
presence of $k_{\max }$.  (We also note that at the critical point, it is
$\tau_{\perp}$ that vanishes in KLS, whereas $\tau_{||}$ vanishes in the present case.)
To retrieve the undriven case, we need only
set $k_{\max }=0$, $\nu _{||}=\nu _{\bot }=\gamma _{\times }=\gamma
_{\bot }=1$, and $\tau _{||}=\tau _{\bot }$. Though this form most likely
has limited validity (e.g., the presence of the singular expression $\left\vert
q_{||}\right\vert $ for all $q_{\bot }$), there is no good theoretical basis
for us to propose additional terms at present. Thus, we will discuss
the salient features associated with this expression.

Recall that, in our notation, $S_{||}\left( k\right) $ is given by Eq.~(\ref%
{PhenoS}) with $q_{||}=k$ and $q_{\bot }=0$, while $S_{\bot }\left( k\right)
$ is obtained when we set $q_{||}=0$ and $q_{\bot }=k$. As in the KLS model, the
\textit{discontinuity singularity }is induced by violation of detailed
balance \cite{JSLC} and displayed through%
\begin{equation*}
\Delta \equiv \frac{\nu _{||}}{\tau _{||}+k_{\max }^{2}}-\frac{\nu _{\bot }}{%
\tau _{\bot }}\neq 0
\end{equation*}%
Considering the points nearest the origin
(for this system size), we have
$S\left( \pi /50,0\right) \simeq 1.32$ and $S\left( 0,\pi /50\right) \simeq
0.78$ (Fig.~\ref{sparsprp}), so that $\Delta \simeq 0.54$. This singularity
is manifested in configuration space by $G\left( \vec{r}\right) $ decaying
as $\left( \Delta /2\pi \right) (y^{2}-x^{2})/(x^{2}+y^{2})^{2}$ at large
distances\cite{LRC}. Although such a power will dominate the ordinary
exponential decay for sufficiently large $r$, the crossover will depend on
the ratio of the amplitudes of the two contributions. In this case, there appears to be
only a small regime ($10\lesssim r\lesssim 20$) where power laws \cite{finiteL},
are found in both $G_{||}$ and $G_{\bot }$. While the data are
consistent with $G_{||}$ being $-\Delta /\left( 2\pi x^{2}\right) $, the
case for $G_{\bot }$ appears to support the power $-3$ instead. If more
extensive simulations bear out this scenario, then it will be a challenge to
understand such anomalous, yet \textquotedblleft generic\textquotedblright\
(in the sense of being present far from critical points)\ singularities.

\begin{figure}[h!]
\centering
\includegraphics[scale=0.7]{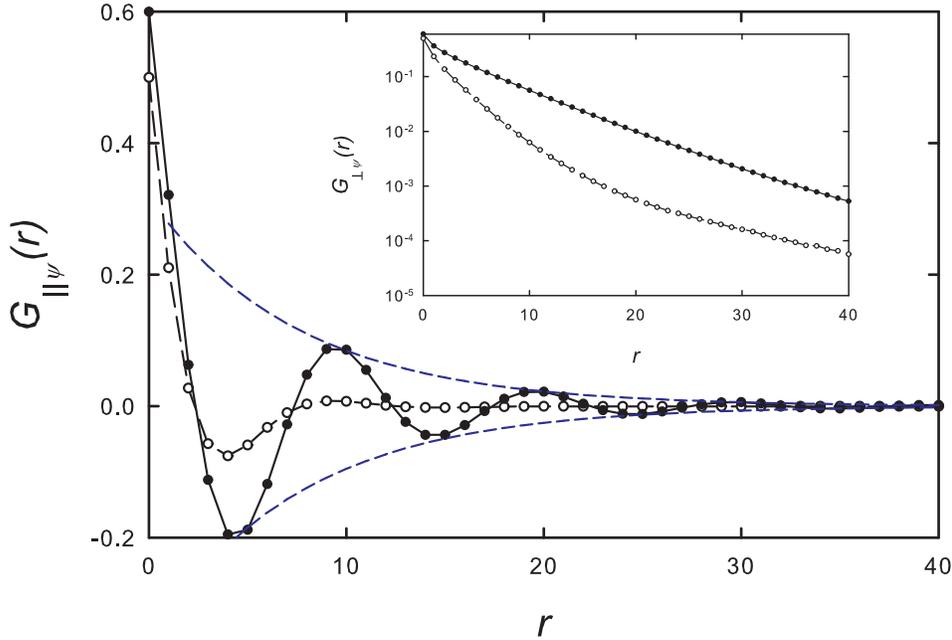}
\caption{(Color online) Main graph: $G_{||}$ for densities 0.5 (open symbols) and 0.6 (filled symbols), for
$p=1$, $a=0$ and $L=100$. The dashed curves are an exponential envelope fit to the
data for density $0.6$.  Inset: $G_{\bot }$ for the same parameters (note semilog scale).}
\label{Ga0p1r5+6}
\end{figure}

Let us return to the more prominent feature, namely, the presence of a
maximum at non-vanishing $q_{||}$, accompanied by the singular $\left\vert
q_{||}\right\vert $ at the origin. Not surprisingly, the main consequence is
oscillations in $G_{||}$, within an exponential envelope. In the $\rho =0.4$
case, $\xi_{||} \ll \lambda $, so that the oscillations implied by $\lambda $ are
not easily seen. These features are however prominent at
larger densities, where $\xi_{||} $ is more comparable to $\lambda $. Meanwhile,
as might be expected, no oscillations are found in $G_{\bot }$.
In Fig.~\ref{Ga0p1r5+6} we show both $G_{||}$ and $G_{\bot }$ for systems with $\rho
=0.5 $ and $0.6$ (again, with $a=0$, p=1 and $L=100$). The rapid increase in $\xi_{||} $
over that for $\rho =0.4$, along with hardly any changes in $\Delta $%
, implies that the power-law decays in $G$ are mostly masked. To be more
confident of the nature of such powers (expected or \textquotedblleft
anomalous\textquotedblright ), we will need to perform longer simulations
(for statistics) of larger systems (to open a larger window between $\xi_{||} $
and $L/2$).

As shown in Fig.~\ref{FFS}, deep in the disordered phase ($\rho=0.4$),
finite-size effects appear to be minimal already for $L=O\left( 100\right) $.
It is clear that, for this density, there is no need to simulate larger systems to
capture the essence of the prominent properties of the driven WRLG.
By contrast, in the next section, we show that the maximum in $S\left(
\vec{k}\right) $ increases with $\rho $, so that, for sufficiently large
densities, the value does not converge as $L\rightarrow \infty $. Instead, $%
S_{||}\left( k_{\max }\right) $ will diverge as the system size, so that $%
S_{||}\left( k_{\max }\right) /L^{2}$ can serve as an order parameter, in
the same spirit as $S\left( \vec{0}\right) /L^{2}$ is a measure of the
(square of the) spontaneous magnetization in the non-conserved Ising
model in two or more dimensions. Along these lines, we mention for comparison, the divergence of $%
S\left( 0,2\pi /L\right) $ or $S\left( 2\pi /L,0\right) $ for the 2D Ising
lattice gas and of $S\left( 0,2\pi /L\right) $ alone for the KLS model. Different from these
examples, here $S$ diverges at an $L$-independent wave-vector $\left(
k_{\max },0\right) $ over a sizable region of the $\rho $-$p$ plane.

\subsection{Discrete mean-field theory and its continuum limit}

In this subsection, we present the simplest theoretical description, as the
first, small step towards a more comprehensive theory. The idea here is to
consider two continuous variables at each discrete site, each corresponding
to the average occupation by the two species: $\rho _{A,B}\left( \vec{x}%
,t\right) $. Next, we postulate the simplest gain/loss terms for the change
in $\rho _{A,B}$ in a single time step. Instead of writing every term here,
let us illustrate with just one example: the loss to $\rho _{A}\left(
\vec{x},t\right) $ due to $A$ particles hopping along $x$ (the $t$ argument is dropped for
simplicity):

\begin{align}
& \frac{\left( 1\!-\!a\right) \left( 1\!+\!p\right) }{6}\rho _{A}\left( x,y\right)
h\left( x\!+\!1,y\right) \left[ 1\!-\!\rho _{B}\left( x\!+\!1,y\!+\!1\right) \right] \left[
1\!-\!\rho _{B}\left( x\!+\!1,y\!-\!1\right) \right] \left[ 1\!-\!\rho _{B}\left(
x\!+\!2,y\right) \right] +  \label{DMF1} \\
& +\frac{\left( 1\!-\!a\right) \left( 1\!-\!p\right) }{6}\rho _{A}\left( x,y\right)
h\left( x\!-\!1,y\right) \left[ 1\!-\!\rho _{B}\left( x\!-\!1,y\!+\!1\right) \right] \left[
1\!-\!\rho _{B}\left( x\!-\!1,y\!-\!1\right) \right] \left[ 1\!-\!\rho _{B}\left(
x\!-\!2,y\right) \right]  \label{DMF2}
\end{align}%
Here, $h\equiv 1\!-\!\rho _{A}\!-\!\rho _{B}$ is the density of holes. The rest of the analysis is
straightforward, though quite tedious. It is clear that such terms would be
exact if we had written the averages of the {\it products} of the appropriate $%
\sigma $s. In a mean-field (MF) approximation, these are replaced by the
appropriate products of the averages, e.g., $\left\langle \sigma \left(
1-\left\vert \sigma \right\vert \right) \right\rangle \rightarrow (\rho
_{A}-\rho _{B})h$.
Beyond the site approximation pursued here, a (generally) more reliable (and complicated)
approach is the pair approximation, which treats nearest-neighbor two-site joint probabilities as the
basic elements \cite{marro}. Although we plan to implement the pair approximation in future work, it seems possible
that only more sophisticated theories will be able to capture the periodic structure observed
under a drive.

Since $\rho _{A}\left( \vec{x},t\right) $ is a conserved density, the gain
and loss terms are necessarily representable in terms of divergences of
currents. Identifying the terms associated with the \textit{net} hops from,
say, $x$ to $x+1$, we find the leading contribution by setting both
densities to be the overall value: $\rho _{A,B}=\rho /2$. This
is the same approach used to find the current-density relationship $J_{KLS}\left(
\rho \right) =\rho \left( 1-\rho \right) $ in the KLS model. The result here is
\begin{equation}
J_{MF}\left( \rho \right) =\left( 1-a\right) p \, \frac{\rho }{2}\left( 1-\rho
\right) \left( 1-\frac{\rho }{2}\right) ^{3}
\label{JMF}
\end{equation}
where the various factors can be readily associated with those in, say, Eq. (%
\ref{DMF1}).

\begin{figure}[h!]
\centering
\includegraphics[scale=0.7]{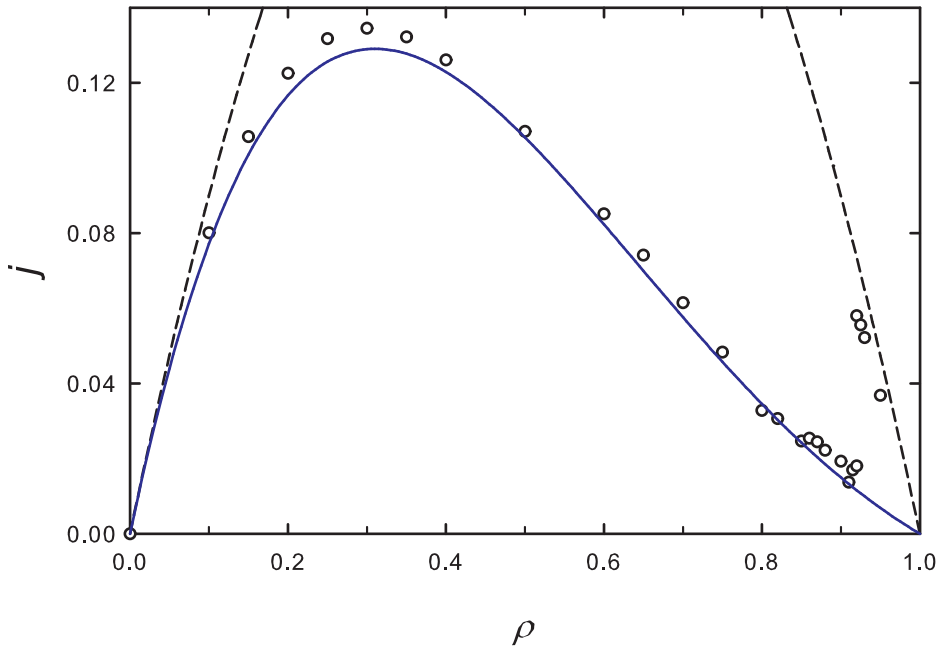}
\caption{(Color online) Current density $j$ versus particle density $\rho$ for drive parameters $p=1$ and $a=0$,
system size $L=200$.  Points: simulation results; solid curve: mean-field prediction, Eq.~(\ref{JMF});
dashed curve: simple mean-field prediction, $j=\rho(1-\rho)\equiv J_{KLS}$.}
\label{CurrentDensity}
\end{figure}

In Fig.~\ref{CurrentDensity}, we show this $J_{MF}\left( \rho
\right) $ (dashed line) and the simulation data (symbols) for a system of size $L=200$
with $a=0,p=1$, as well as $J_{KLS}\left( \rho \right) $ (dotted
line) for comparison. It is remarkable how the properties of a complicated
stochastic system are mostly captured (i.e., $\rho \lesssim 0.9$) by such a
simple minded approach. Near the $\rho =1$ limit, we note that the data
jump to follow another \textquotedblleft branch\textquotedblright\ of the $%
J $-$\rho $ relationship, which is much closer to $J_{KLS}\left( \rho
\right) $. This seemingly paradoxical result can also be understood, as
follows. The jumps in the data are associated with
\textquotedblleft merging\textquotedblright\ transitions, where many strips
merge into fewer. Eventually, the system settles into a single
strip (for each species) configuration. In this limit, it is natural to
regard the holes as \textquotedblleft particles\textquotedblright\ and
particles as \textquotedblleft holes.\textquotedblright\ The idea is that
only some of these \textquotedblleft particles\textquotedblright\ are bound
to interfaces which separate the $A$- and $B$-rich domains. Roughly
speaking, $\rho L^{2}-2L$ of them remain mobile and are driven through the
large domains of \textquotedblleft holes,\textquotedblright\ interacting in
a minimal way with the ones bound to the interfaces. Thus, we can apply $%
J_{KLS}$ to this system, with the density of \textquotedblleft
particles\textquotedblright\ being $\rho -2/L$ and (mobile)
\textquotedblleft holes\textquotedblright\ being $1-\left( \rho -2/L\right) $%
. In the large $L$ limit, such an expression reduces precisely to $\rho
\left( 1-\rho \right) $. Without showing it explicitly, the agreement
between data and $\left( \rho -2/L\right) \left( 1-\rho +2/L\right) $, with $%
L=200$, is excellent.
The reason the current associated with single-stripe configurations
lies much closer to the simple KLS prediction is that in this case
the density of {\it mobile} vacancies is much nearer $1-\rho$. 

In Fig.~\ref{CurrentDensity}, there is a range of densities
(approximately $0.86 \leq \rho \leq 0.87$) in which both multistripe
and single-stripe configurations exist.  Although the
former appear to be metastable, precise determination of the associated
lifetimes is left for future study.  On a lattice of $L \times L$ sites,
the maximum possible density is $1 - L/2$.  Configurations with this
density are frozen, since there are then {\it no} mobile vacancies.
Outside this limit, however, we do not observe jamming in the sense
found in the driven NNE model with NN hopping dynamics \cite{drnne}.

The next natural step in the attempt to formulate a theoretical approach is
to write a continuum approximation for the discrete MF, while keeping the
lowest few orders in the expansion of the displaced densities, e.g., $\rho
\left( x+1\right) \rightarrow \rho \left( x\right) +\varepsilon \partial
_{x}\rho +\left( \varepsilon \partial _{x}\right) ^{2}\rho /2+...$. The final
step is to limit our present study to the homogenous phase and consider the
lowest orders in the expansion of the densities
\begin{equation*}
\rho _{A,B}\left( \vec{x},t\right) =\frac{\rho }{2}+\phi _{A,B}\left( \vec{x}%
,t\right) +...
\end{equation*}%
In this manner, we obtain expressions for $\partial _{t}\phi _{A,B}\left(
\vec{x},t\right) $. If we keep only the order linear in $\phi $, we will
find
\begin{equation}
\partial _{t}\left(
\begin{array}{c}
\phi _{A} \\
\phi _{B}%
\end{array}%
\right) =\left(
\begin{array}{cc}
D & \tilde{D} \\
\tilde{D} & D%
\end{array}%
\right) \left(
\begin{array}{c}
\phi _{A} \\
\phi _{B}%
\end{array}%
\right)   \label{linear}
\end{equation}%
where $D$ and $\tilde{D}$ are differential operators. Note that we have
taken account of the symmetry under $A\Leftrightarrow B$ to write this form.
Clearly, the sum and difference of the $\phi $\ fields diagonalize this
matrix. There should be no confusion if we again use the notation $%
\psi \left( \vec{x},t\right) $ for $\phi _{A}-\phi _{B}$. In keeping with
the discussions above, we will focus on the $\psi $ equation only, though
the sum, $m\left( \vec{x},t\right) \equiv \phi _{A}+\phi _{B}$, will play a
role at the nonlinear level. The result of a tedious computation is, apart
from the details of the coefficients, to be expected from the fact that $%
\psi $ is a conserved density. Thus, we find $\partial _{t}\psi $ to be the
divergence of a current density, obeying the appropriate space-time
symmetries:%
\begin{equation*}
\partial _{t}\psi =\left[ \left( E_{1}\partial _{x}+E_{3}\partial
_{x}^{3}...\right) +\left( D_{||}\partial _{x}^{2}+D_{\bot }\partial
_{y}^{2}+...\right) \right] \psi
\end{equation*}%
For clarity, we have kept the parts odd in $\partial _{x}$ (which must be
absent when $p=0$) separate from those even in $\partial _{x,y}$ (which
represent the diffusive parts of this equation). Thus in Fourier space $%
\left( \vec{q},\omega \right) $, the former set appears with $i$ and can be
seen as the real part of a dispersion relation, $\omega \left( \vec{q}%
\right) $, which plays the role of driven transport. The imaginary part
plays the role of damping, providing the sign is appropriate. From these, we
have also allowed for anisotropic diffusion ($D_{||}\neq D_{\bot }$), as $p,a
$ are not necessarily $0,1/4$. Of course, the various coefficients are real
functions of $\rho $, $p$ and $a$, the only control parameters here. At this
level, all are expected to be \textit{analytic}, so that we expect the $E$'s
and $D$'s to be odd and even functions, respectively, of the drive $p$.
Naturally, the next step is to explore (linear) instability, by checking if
any aspect of the damping vanishes. The only such occurrence is $D_{||}$
vanishing at $\rho =1/6$, independent of $p$, corresponding to the
the unmixing transition in the equilibrium WRLG.
Needless to say, this
instability is of little relevance to our driven lattice gas. But, as a
mean-field-like result, it may serve as a starting point for serious analysis.

To gain some insight into the structure factors and correlations, we must
add noise terms to form a set of Langevin equations for the density fields,
though these terms cannot be derived from the discrete mean-field equations
above. Thus we write,
\begin{equation}
\partial _{t}\psi =\left[ \left( E_{1}\partial _{x}+E_{3}\partial
_{x}^{3}...\right) +\left( D_{||}\partial _{x}^{2}+D_{\bot }\partial
_{y}^{2}+...\right) \right] \psi +\eta \left( \vec{x},t\right).
\label{Langevin-psi}
\end{equation}%
To account for the conservation law and possible anisotropy, we will assume
that $\eta $ is delta-correlated, conserved, Gaussian noise, with zero mean
and
\begin{equation*}
\left\langle \eta \left( \vec{x},t\right) \eta \left( \vec{x}^{\prime
},t^{\prime }\right) \right\rangle =-\left[ N_{||}\partial _{x}^{2}+N_{\bot
}\partial _{y}^{2}\right] \delta \left( \vec{x}-\vec{x}^{\prime }\right)
\delta \left( t-t^{\prime }\right) .
\end{equation*}%
Here again we have allowed for anisotropy, by writing two different
coefficients. As a reminder, in the undriven limit, our dynamics satisfy
detailed balance, which implies $N_{||}/D_{||}=N_{\bot }/D_{\bot }$.
Following standard routes pioneered by Martin-Siggia-Rose \cite{MSR},
Janssen \cite{HKJ}, and de Dominicis \cite{CdD}, we write the dynamic
functional $\mathcal{J}\left[ \psi ,m;\tilde{\psi},\tilde{m}\right] $ for
both the original density fields and the associated response fields $\tilde{%
\psi},\tilde{m}$. At the quadratic level, corresponding to the linear Langevin
example above, the $\psi $ and $m$ sectors decouple.
The propagators and correlators (e.g., in the $\psi $ sector, the $\psi \tilde{\psi}$ and
$\tilde{\psi}\tilde{\psi}$ terms) in Fourier space are, respectively,
$i\left[ \omega -\omega \left( \vec{q}\right) \right] +D_{||}q_{x}^{2}+D_{\bot }q_{y}^{2}+O\left( q^{4}\right)$
and $N_{||}q_{x}^{2}+N_{\bot }q_{y}^{2}$.
The most general $O\left( q^{4}\right) $ terms consistent with symmetry,
anisotropy, and analyticity are just $\Gamma _{||}q_{x}^{4}$, $\Gamma
_{\times }q_{x}^{2}q_{y}^{2}$ and $\Gamma _{\pm }q_{y}^{4}$ . Thus, at this
tree level, the equal-time correlation $G_{tree}\left( \vec{r},0\right) $
can be found easily; its transform reads,%
\begin{equation*}
S_{tree}\left( \vec{q}\right) =\frac{N_{||}q_{x}^{2}+N_{\bot }q_{y}^{2}}{%
D_{||}q_{x}^{2}+D_{\bot }q_{y}^{2}+O\left( q^{4}\right) }
\end{equation*}%
While this approach may appear to be reasonable for describing the
fluctuations of the homogeneous state, clearly, it fails to predict the most
prominent feature: presence of a $\left\vert q_{x}\right\vert $ term and the
associated maximum at $k_{\max }$. We are exploring  non-linear terms in the
Langevin equation (beyond quadratic in $\mathcal{J}$) and whether a simple
perturbative approach can generate such a kink singularity.
We conjecture that at one-loop level, a term
$\propto \left\vert q_{||}\right\vert ^{3}$ will emerge in the denominator, thereby
justifying the form in Eq. (5).
Work is in
progress to examine a systematic approach, applying the Doi-Peliti formalism \cite{Doi,Peliti,Pert,DPrev}
to the defining microscopic dynamics to derive (instead of postulating) the
appropriate functional. It is our hope that results from such an
analysis will provide a better description of the phenomenological form
above. The success of stochastic field theory in the past is, of course, the
systematic study of universal properties in critical phenomena. Assuming
that the transition from the homogeneous to the inhomogeneous state(s) is
continuous, we believe this line of pursuit will lead to novel fixed
point(s) and universality class(es).

\section{Inhomogeneous states: Phase segregation, multiple stripes,
correlations, and interfacial properties}

In this section we report simulation results on the ordered phase of the
driven WRLG. In equilibrium, for $\rho_A = \rho_B = \rho/2$, the system
exhibits separation into A- and B-rich phases above a certain critical
density. We therefore begin by looking for signs of phase separation under a
drive.

\subsection{Emergence of striped phases under a drive}

As shown in the preceding section, under a drive, the disordered system exhibits a
preferred wavelength $\lambda$ for charge density oscillations.
We find that, for densities $\rho$ near the onset of phase
separation, a striped pattern of wavelength $\lambda$ appears, as is evident
in configuration snapshots as well as quantitative measures such as the
structure factor. With increasing density, $\lambda$ first decreases,
attaining a minimum $\lambda_{min}$ for $\rho \simeq 0.65-0.70$, and then
grows, approaching the system size $L_x$, as $\rho \to \rho_{max} = 1 -
2/L_x $. (Recall that at least $2L_y$ vacancies are required to satisfy the
prohibition of NN A-B pairs.) We expect that in the limit $L_x, L_y \to
\infty$, with fixed $L_y/L_x$ and $\rho < 1$, the wavelength $\lambda$
converges to a size-independent function of the density and the drive
parameters.

In equilibrium, of course, maximum entropy implies the minimum possible number of interfaces,
so that $\lambda$ is formally infinite for zero drive. (There is, of course,
no striped pattern in equilibrium.) Consistent with this, $\lambda$ grows
systematically as the drive $p \to 0$. For $p = 1$, 1/2, 1/3, 1/4 and 1/8,
we find $\lambda_{min}$ = 10, 15, 18, 22 and 36 respectively; these data
follow $\lambda \simeq C_1 + C_2/p$ where $C_1$ and $C_2$ are constants.

In contrast to the KLS model, stripes of distinct phases orient \textit{%
perpendicular} to the drive, and have \textit{rough} interfaces. For a wide
range of densities, single-stripe configurations (i.e., $\lambda=L_x$) are
\textit{unstable}, and the steady state exhibits multiple stripes; for the
system sizes investigated, single-stripe configurations do appear to be
stable for higher densities (specifically, for $\rho \geq 0.83$ when $a=0$
and $p=1/3$, and for $\rho \geq 0.93$, for $a=0$ and $p=1$). These
observations are illustrated for $a=0$ and $p=1/3$, in Figs.~6a
and \ref{cfg65}. For the densities shown, A- and B-rich regions are evident,
organized into stripes oriented perpendicular to the drive. At the highest
density shown, $\rho=0.85$, each phase occupies a single stripe, a situation
that persists for systems of size $L \leq 400$. The configurations shown in
Figs.~6-\ref{cfg68} were obtained using single-stripe initial conditions
(perpendicular to the drive), thus demonstrating instability toward
formation of multiple stripes for $\rho \leq 0.82$.

We note that, for $\rho > \rho_c$ and a nonzero drive, initial
configurations with A- and B-regions separated by a boundary \textit{parallel%
} to the drive rapidly reorganize so that resulting the stripe or stripes
are perpendicular to it. After a short time ($\sim 10^3$ MC steps) the
initially flat interface develops waves which grow steadily in
amplitude. When they reach a height $\sim L/2$, they reconnect via the
periodic boundaries, forming stripes perpendicular to the drive. The latter
are initially quite irregular, but (for sufficiently high densities) rapidly
become ordered, so that the system is separated into clear bulk and
interfacial regions.

\begin{figure}[h]
\label{cfgsp1} \center
\subfigure[]{\includegraphics[scale=0.49]{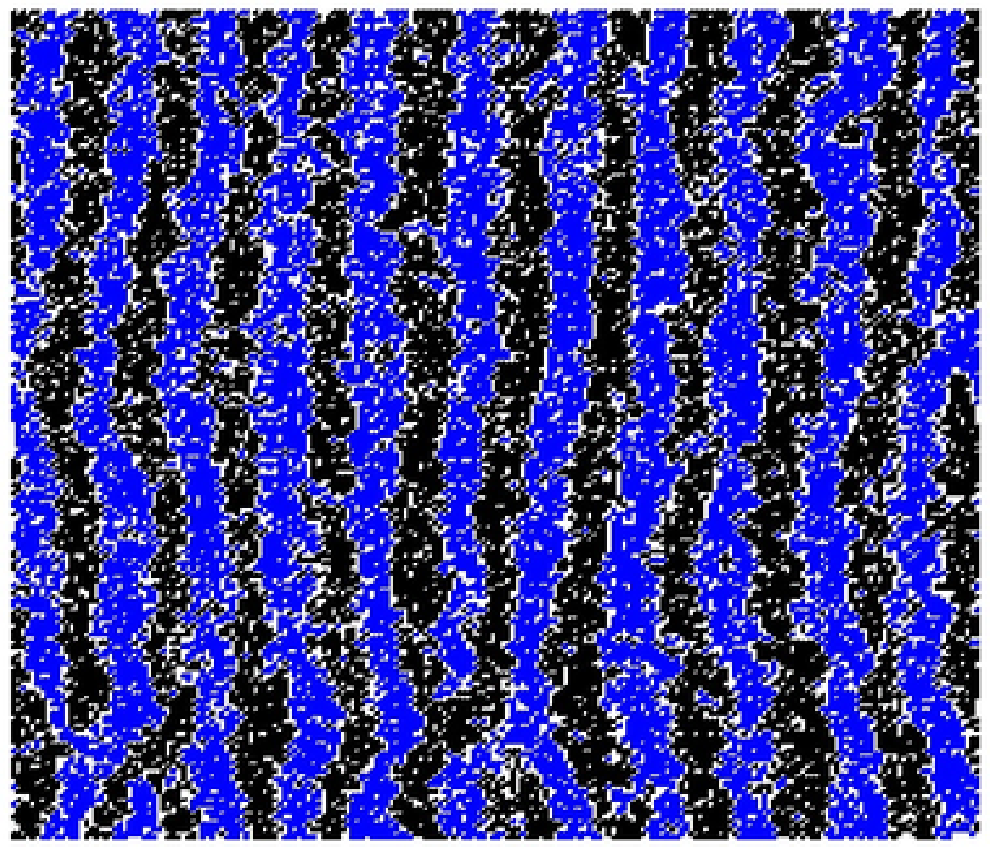}} \qquad %
\subfigure[]{\includegraphics[scale=0.49]{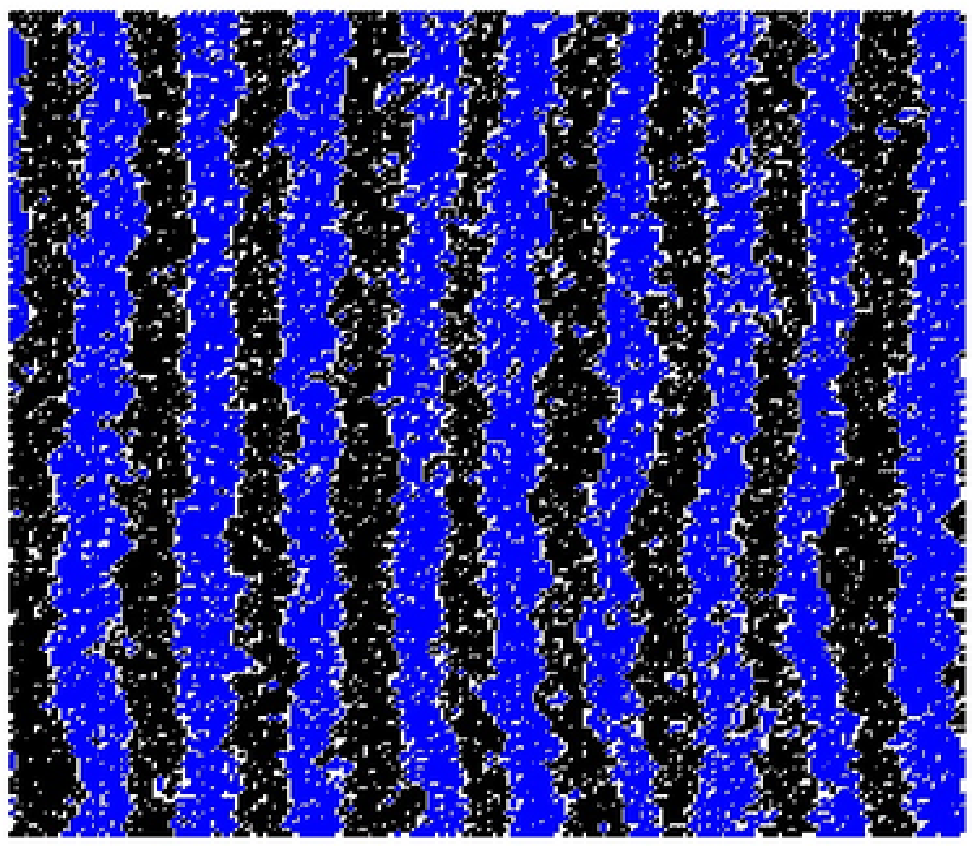}} \qquad %
\subfigure[]{\includegraphics[scale=0.49]{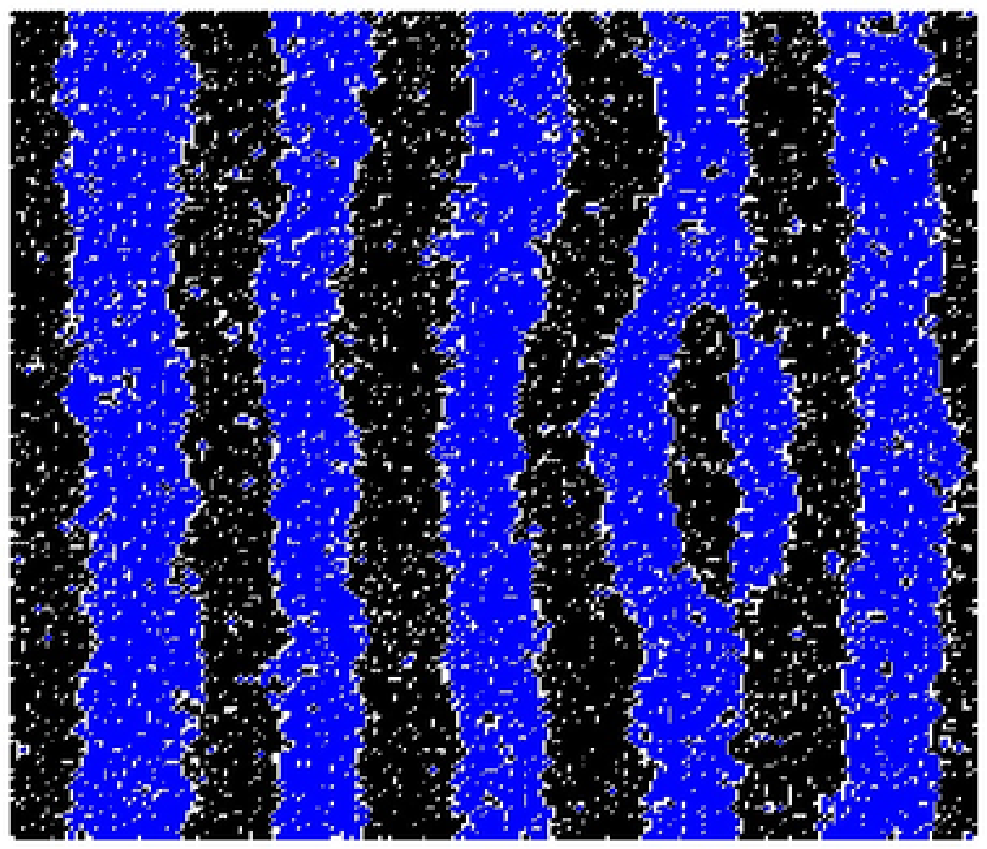}} \qquad %
\subfigure[]{\includegraphics[scale=0.49]{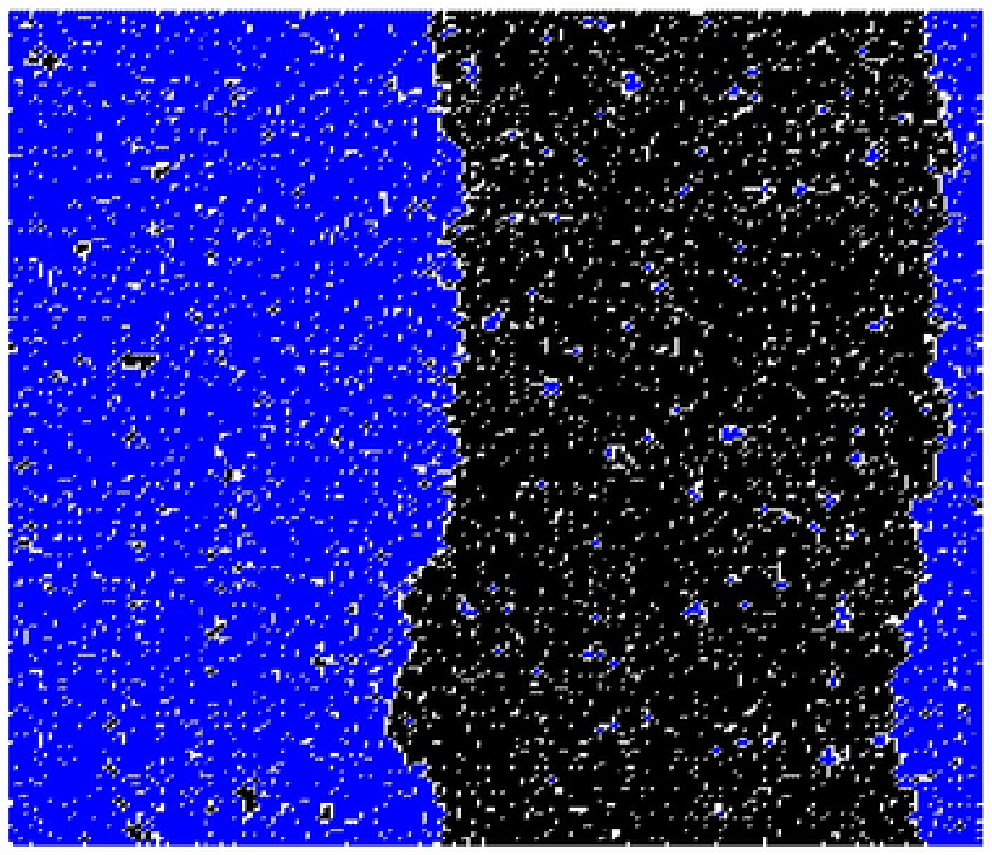}}
\caption{(Color online) Typical configurations of driven Widom-Rowlinson lattice gas,
system size $L=200$, parameters $a=0$ and $p=1/3$, with the drive oriented
toward the right. Points of differing color denote particles of species A and B,
respectively; white points denote vacant sites. Particle densities $\protect%
\rho = 0.7$ (a), 0.75 (b), 0.80 (c), 0.85 (d). }
\end{figure}

\begin{figure}[h!]
\centering
\includegraphics[scale=0.7]{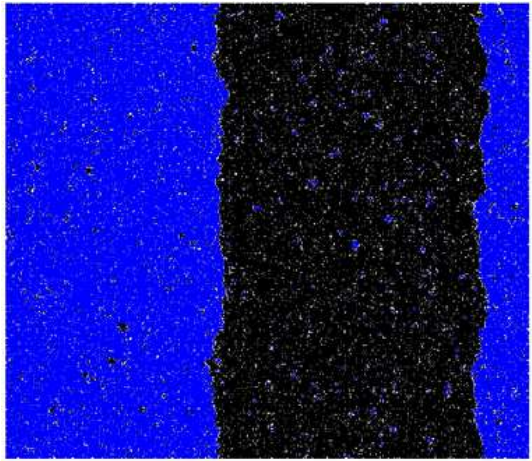}
\caption{(Color online) Configuration for $L=400$, $\protect\rho = 0.85$, $a=0$ and $p=1/3$.}
\label{cfg65}
\end{figure}

While experience with the KLS model might lead one to expect interfaces
parallel to the drive, it appears that without attractive interactions there
is no mechanism for stabilizing and smoothing such an interface in the
driven WRLG. On the other hand, sequences of the form A0B (where 0
represents a vacant site) along the drive represent barriers, behind which
particles may form a queue (and analogously for B0A sequences). Formation of
queues (together with the prohibition of A-B NN pairs) may stabilize stripes
perpendicular to the drive. Why many stripes of the same phase form, instead
of just one, is unclear. Figure 6c provides a hint of the mechanism: a
fluctuation in the boundary has allowed B particles to invade an A stripe,
leading to formation of a new stripe. Multi-stripe configurations may contain
defects, as is evident in Figs.~6a and \ref{cfg68}.

\begin{figure}[h!]
\centering
\includegraphics[scale=0.7]{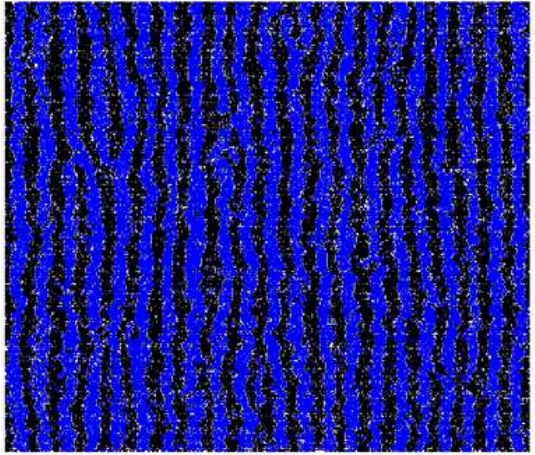}
\caption{(Color online) Configuration for $L=400$, $\protect\rho = 0.7$, $a=0$ and $p=1/3$.}
\label{cfg68}
\end{figure}

\subsubsection{The preferred wavelength}

\begin{figure}[h!]
\centering
\includegraphics[scale=0.6]{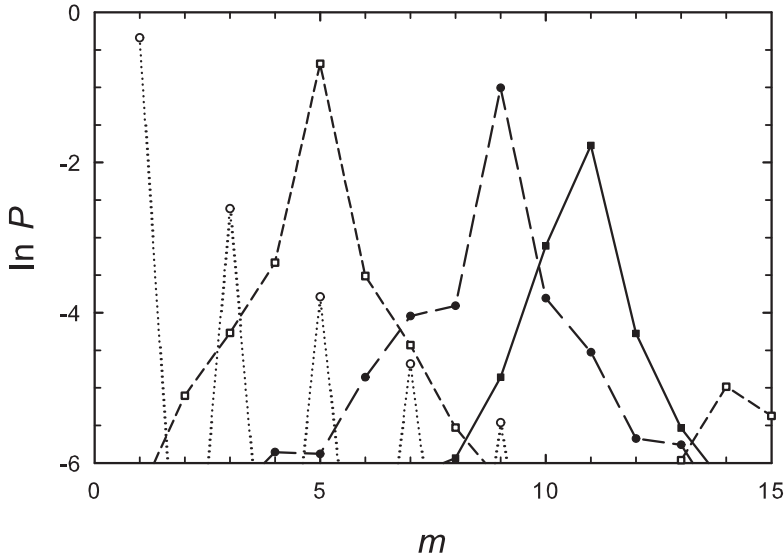}
\caption{Power spectra of composition variation $\protect\psi_{||,i}$ along
the drive direction for $L=200$, $p=1/3$, $a=0$, and $\protect\rho = 0.85$
(open circles), 0.80 (open squares), 0.75 (filled circles) and 0.70 (filled
squares).}
\label{pwrsp}
\end{figure}

\begin{figure}[h!]
\centering
\includegraphics[scale=1.2]{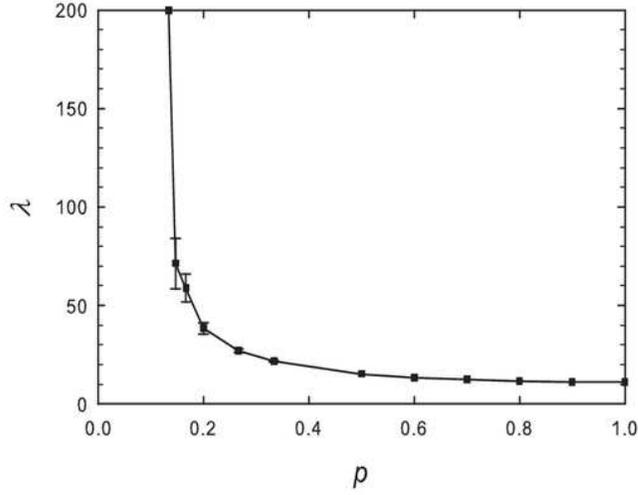}
\caption{Preferred wavelength $\protect\lambda$ versus drive parameter $p$
for $L=200$, density $\protect\rho=0.75$ and hopping parameter $a=0$. Since
the system size is 200, the preferred wavelength at the lowest density shown
may well exceed 200; studies of larger systems would be necessary to verify
its value.}
\label{lambvp}
\end{figure}

\begin{figure}[h!]
\centering
\includegraphics[scale=0.6]{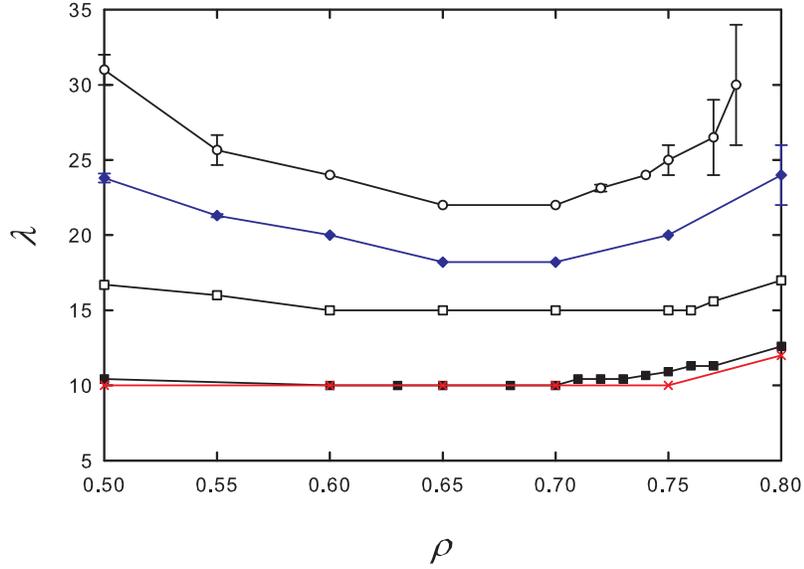}
\caption{(Color online) Preferred wavelength $\protect\lambda$ versus density $\protect\rho$
for drive parameters (upper to lower) $p=1/4$, 1/3, 1/2 and p=1; hopping
parameter $a=0$ in all cases. System sizes: $L=240$ ($p=1/4$), 200 ($p=1/3$
and 1/2), 240 ($p=1$: filled squares) and 120 ($p=1$: $\times$ ). Lines are
guides to the eye.}
\label{lambvrho}
\end{figure}

As noted above, for a nonzero drive, the structure factor $S_{||,k}$ shows a
distinct maximum for a certain wavenumber, $k = 2 \pi m/L$, corresponding to
the preferred wavelength $\lambda = L/m$. Power spectra of the columnwise
charge density $\psi_{||,i}$ are particularly useful in determining $\lambda$
(see Fig.~\ref{pwrsp}). The latter is a decreasing function of drive at
fixed density, as shown in Fig.~\ref{lambvp}. Since the density of vacancies
$1-\rho$ is complementary to the particle density, and since each stripe
requires at least $2L$ vacancies for the interfaces, the number of stripes
should decrease (and $\lambda$ increase) as the density approaches unity, as
verified in Fig.~\ref{lambvrho}. This figure also shows that the dependence
of $\lambda$ on $\rho$ for different drive strengths is qualitatively
similar, and that the dependence of $\lambda$ on system size is weak.

Our results for $\lambda$ derive from initial configurations (ICs) using
both single and multiple stripes. For densities in the range $0.5 \leq \rho
< 0.8$, the different ICs yield consistent results for the stationary value
of $\lambda$. For $\rho \geq 0.8$, however, the steady-state wavelength $%
\lambda_s$ tends to remain the same as the initial value, $\lambda_0$, for
the duration of the simulations, over a wide range of $\lambda_0$, hampering
a precise determination of $\lambda_s$. For $p=1/3$ and $\rho=0.8$, for example, we find $%
\lambda_s = \lambda_0$ for initial wavelengths in the range 22-26. For $%
p=1/4 $ and $\rho = 0.8$ the range of stable values broadens to 30-50.
(While it remains possible that the uncertainty in $\lambda$ would be
smaller, using longer studies, it appears that the evolution of the number
of stripes is very slow at densities $\geq 0.8$.)

\subsection{Local ordering}

At intermediate densities, the driven system exhibits \textit{local}
ordering into A- and B-rich stripes, without global ordering of the pattern;
an example is shown in Fig.~\ref{cfg68}. To quantify local ordering, we
consider a function that essentially projects regions of size $\lambda$ onto
a patch of an ideal pattern. Consider a rectangle of $\lambda \times \ell$
sites, i.e., $\mathcal{R} \equiv \{1 \leq x \leq \lambda\} \times \{1 \leq y
\leq \ell \}$. Let $Q_1$ be the excess number of A particles over B
particles in the left half ($1 \leq x \leq \lambda/2$) and $Q_2$ the excess
number in the right half; define

\begin{equation}
\psi_{\mathcal{R}} = \frac{Q_1 - Q_2}{\lambda \ell} \, .  \label{defpsi}
\end{equation}

\begin{figure}[h!]
\centering
\includegraphics[scale=0.6]{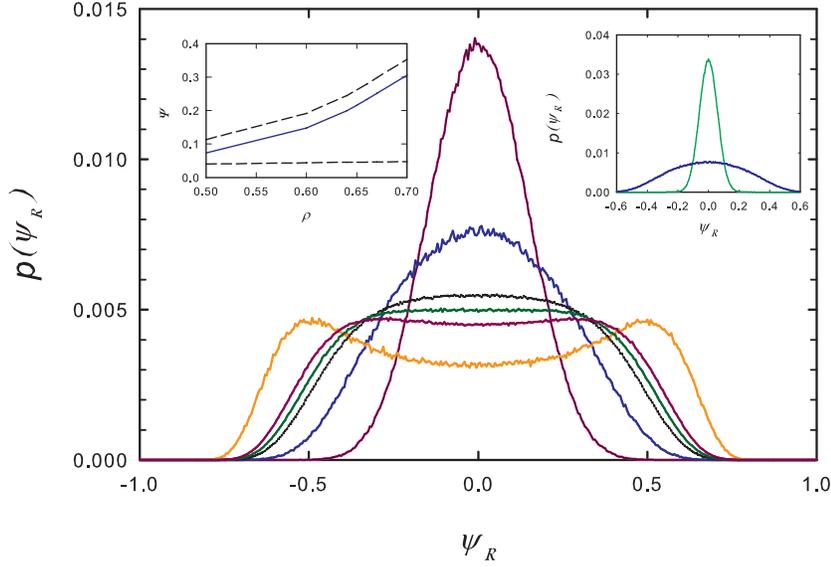}
\caption{(Color online) Probability densities $p(\protect\psi_{\mathcal{R}})$ of the local
order (using $\protect\lambda=10$ and $\ell = 20$), for densities (upper to
lower at center) 0.5, 0.6, 0.64, 0.65, 0.66 and 0.7. Drive parameters $p=1$
and $a=0$, system size $L=100$. Right inset: $p(\protect\psi_{\mathcal{R}})$
for $\protect\rho=0.6$ as in main graph (broad curve), compared with $p_r \,(%
\protect\psi_{\mathcal{R}})$, for randomly generated configurations at the
same density (central curve). Right inset: $\langle | \protect\psi_{\mathcal{%
R}} | \rangle$ (upper dashed curve), $\langle | \protect\psi_{\mathcal{R}} |
\rangle_r$ (lower dashed curve), and their difference, $\Psi$ (central
curve), versus density.}
\label{psinewhst}
\end{figure}

\noindent This quantity takes nonzero values when the region is centered on
a stripe of width $\lambda$, but will be close to zero in random
configurations, or in regions occupied by a single species. Thus a
convenient measure of local ordering is $\Psi = \langle | \psi_{\mathcal{R}}
| \rangle - \langle | \psi_{\mathcal{R}} | \rangle_r$, where the first
average is over configurations in the stationary state at a given density,
and the second is over configurations (of the same density) generated by
inserting A and B particles, with equal probabilities, at random into $%
\mathcal{R}$, subject to the prohibition against A-B nearest-neighbor pairs.
The probability density $p(\psi)$ changes from unimodal to bimodal at a
certain density, marking the growth of local order. In Fig.~\ref{psinewhst},
for $p=1$, the transition from unimodal to bimodal occurs near $\rho = 0.65$%
. This does not imply, of course, that there is no local ordering below this
density. The behavior of $\Psi$ is quite smooth over the range of densities
of interest (see right inset of Fig.~\ref{psinewhst}), suggesting that there
is no phase transition associated with local ordering. For weaker drives,
the transition from a unimodal to a bimodal distribution occurs at a similar
density: $\rho=0.66$ for $p=1/2$, and $\rho=0.63$ for $p=1/4$. Thus, under a
drive, short-range order appears at a density slightly above the equilibrium
critical density, $\rho_c = 0.618(1)$.

\subsection{Order parameter and phase boundary}

\begin{figure}[h!]
\centering
\includegraphics[scale=0.6]{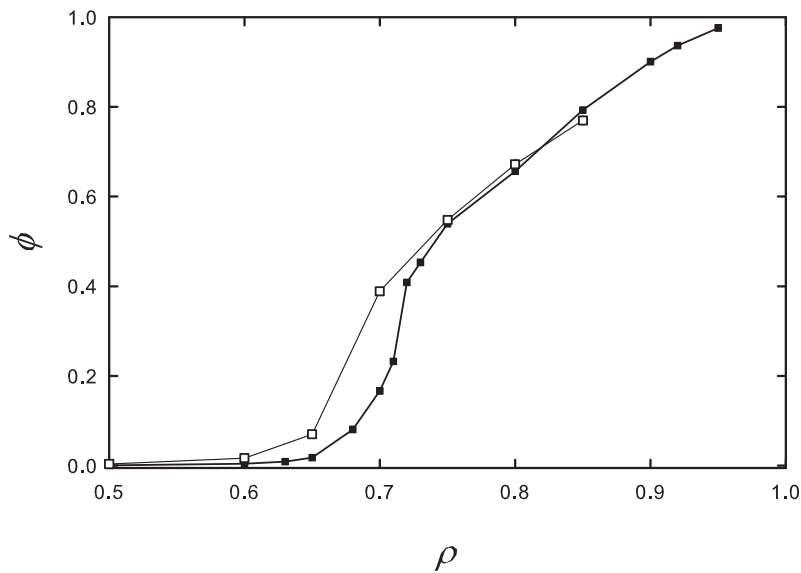}
\caption{Order parameter versus density for $p=1$, $a=0$ and $L=240$ (filled
symbols) and 120 (open symbols). Lines are guides to the eye; error bars
smaller than symbols.}
\label{opvrhop1}
\end{figure}

Given the evidence of phase separation discussed above, in the form of A-
and B-rich stripes oriented perpendicular to the drive, we turn to a more
quantitative discussion, which requires definition of an order parameter.
The ordered phase is characterized by charge-density oscillations along the
drive, i.e., a correlation function $C_{||}(r) \simeq A \exp[iq^*r]$ and an
associated structure factor $S_{||,q}$ with a maximum (at $q=q^*$)
proportional to $(LA)^2$. We therefore define the order parameter $\phi
= S_{||,q*}/L^2$.
The order parameter is plotted versus density in Fig.~\ref{opvrhop1}
for $p=1$ and $a=0$; the plot strongly suggests that there is a continuous
phase transition at a density near 0.72.

To obtain a more precise estimate of the density $\rho_c$ marking the onset of global
order, we perform a finite-size analysis of the order parameter $\phi$. For
fixed $p$ and $\rho$ (and $a=0$), we determine $\phi$ for a series of system
sizes $L$. In the disordered phase, we expect $\phi$ to decrease rapidly
with $L$, whereas in the ordered phase it must approach a nonzero limiting
value as $L \to \infty$. A phase transition is marked by slow decay of the
order parameter with $L$, typically a power law, $\phi \sim L^{-\beta/\nu_{||}}$.
Thus in a plot of $\ln \phi$ versus $\ln L$ for fixed density $\rho$, we
interpret upward (downward) curvature as a signal of the ordered
(disordered) phase.

\begin{figure}[h!]
\centering
\includegraphics[scale=0.6]{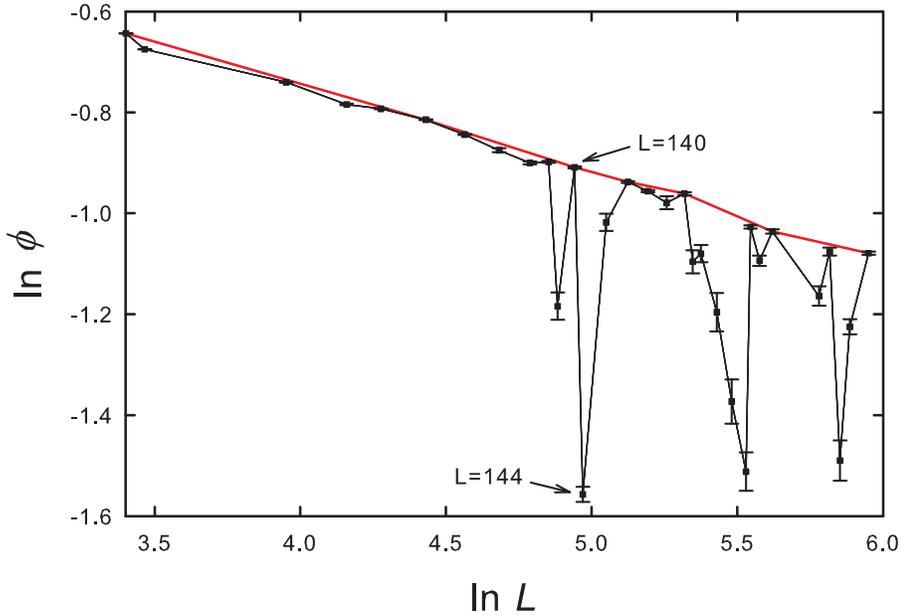}
\caption{Order parameter versus system size for $p=1$, $a=0$ and $\rho = 0.74$.
Values for system sizes 140 and 144 are highlighted for comparison.  The bold line
represents the envelope.
}
\label{phivLp1r74}
\end{figure}

Estimating $\rho_c$ is complicated by the fact that as one varies the
density, the preferred wavelength changes, as shown in Fig.~\ref{lambvrho}.
We obtain more orderly patterns (higher values of $\phi$) using system sizes
$L$ that are integer multiples of the preferred wavelength. (Note that the latter
need not be an integer. For $p=1$ and $\rho=0.74$, for example,
the preferred wavelength is 10.63.) A well ordered pattern consists of an integer number of
wavelengths, which is only possible for
specific system sizes.  This leads to an irregular variation of $\phi$ with $L$,
as shown in Fig.~\ref{phivLp1r74}.  Small changes in $L$ (for example, from 140 to 144) can yield large
(and reproducible) changes in $\phi$.  The various points on the graph of $\phi$ versus $L$
can nevertheless be bounded from above by a smooth ``envelope."  The points falling on or near
the envelope represent the most ordered cases;
we therefore apply the curvature
criterion described above to the envelope.  For the data shown in Fig.~\ref{phivLp1r74},
the $\lambda$ values associated with points along the envelope fall in the range 10.50 - 10.77,
with a mean of 10.63(5), providing an estimate of the preferred wavelength.  We also note that
in this case, the envelope is fairly well described by a power law, with an exponent
$\beta/\nu_{||} = 0.173(5)$.

We adopt the following procedure to determine $\phi$ as a function of $L$:

\begin{enumerate}

\item Perform a survey of L values, to estimate the preferred wavelength $\lambda$.

\item Determine $\phi$ for a series of system sizes $L$ such that there is an integer $n$ with $L/n$ falling
near $\lambda$.

\item If necessary, refine the estimate for the preferred wavelength and return to 2.

\end{enumerate}

\begin{figure}[h!]
\centering
\includegraphics[scale=0.6]{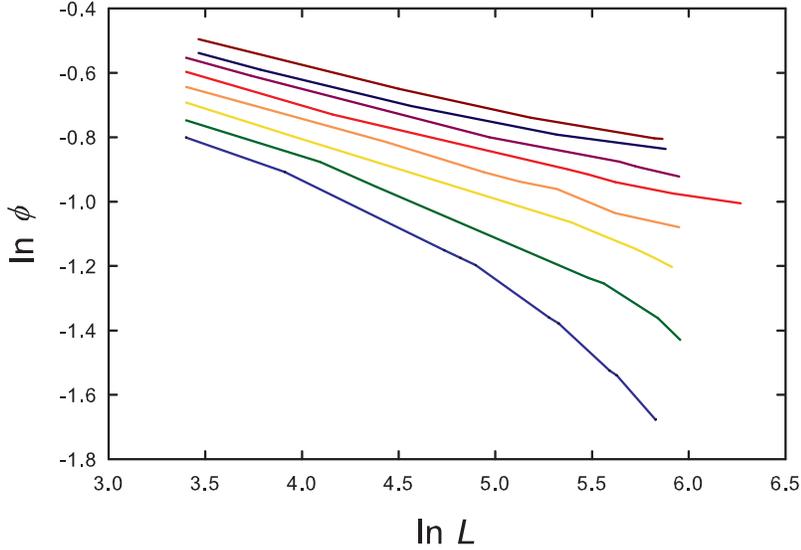}
\caption{(Color online) Envelope of order parameter for $p=1$, $a=0$ and (lower to upper) $\rho = 0.71$,
0.72,..., 0.78.
}
\label{envsp1}
\end{figure}

Varying the density from 0.71 to 0.78 (for fixed $p=1$ and $a=0$), we find the set of envelopes
shown in Fig.~\ref{envsp1}.  For $\rho \leq 0.73$ the envelopes curve downward, while for
$0.74 \leq \rho  \leq 0.76$ there is no clear curvature; for $\rho \geq 0.77$ the data appear
to curve upward.  Thus the onset of global order occurs at a density close to 0.77.
For densities in the range 0.74-0.76, the order parameter appears to decay as a power law
of the system size, with an exponent $\beta/\nu_{||}$ varying between about 0.173 and 0.143.
For a small drive ($p=1/8$, $a=0$), a similar analysis shows that the transition occurs
at $\rho \simeq 0.73$.
Determining the precise value of $\rho_c$ as a function of the drive parameters,
and verifying the existence of a ``critical
phase" in which the order parameter scales as an inverse power of $L$, will require
extensive studies of larger systems, a task we defer to future work.

\subsection{Correlation functions}

\begin{figure}[h!]
\centering
\includegraphics[scale=0.6]{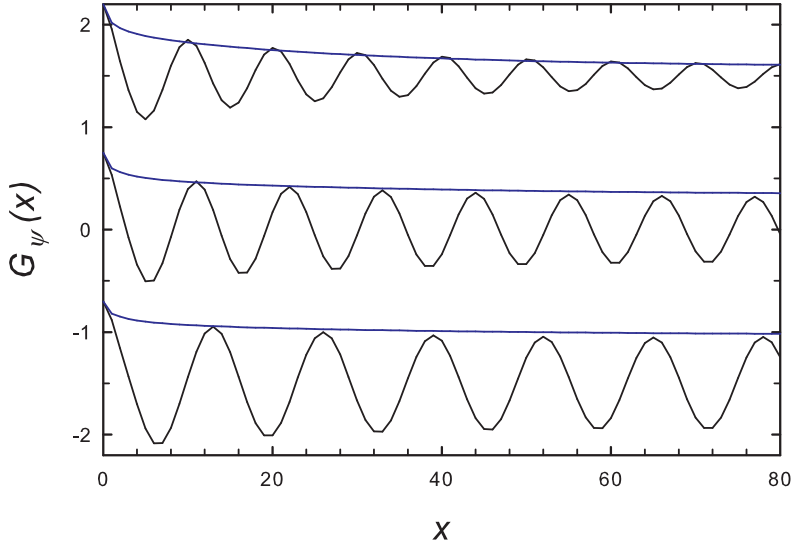}
\caption{(Color online) Charge-charge correlation functions $G_{\protect\psi ||} (x)$,
along drive (oscillating), and $G_{\protect\psi \perp} (x)$, perpendicular
to drive (decaying monotonically), for $p=1$, $a=0$ and (upper to lower)
density $\protect\rho = 0.7$, 0.75 and 0.8. The curves for $\protect\rho=0.7$
(0.8) have been shifted upward (downward) by 1.5 for legibility.}
\label{gpsi}
\end{figure}

\begin{figure}[h!]
\centering
\includegraphics[scale=0.6]{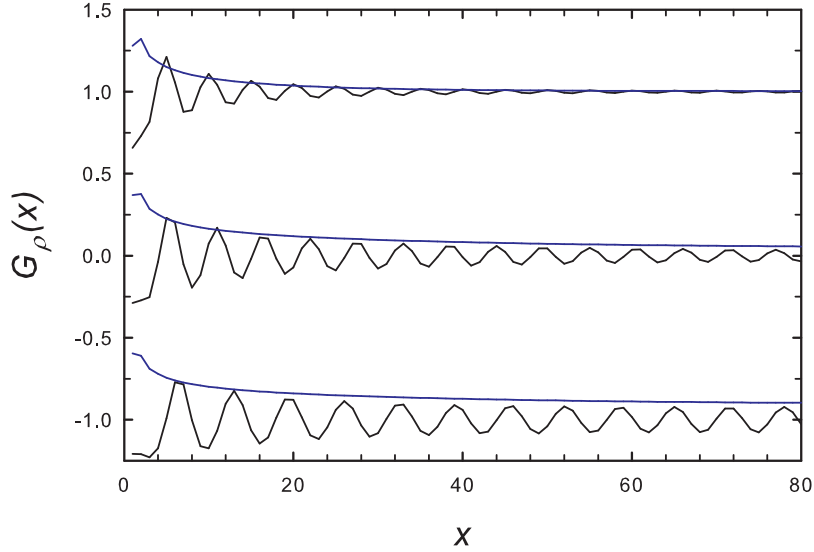}
\caption{(Color online) Density-density correlation functions $G_{\protect\rho ||} (x)$,
along drive (oscillating), and $G_{\protect\rho \perp} (x)$, perpendicular
to drive (decaying monotonically for $x \geq 2$), for parameters as in Fig.
\protect\ref{gpsi}. The correlation functions are multiplied by 10, and
curves for $\protect\rho=0.7$ (0.8) have been shifted upward (downward) by
1, for legibility.}
\label{grho}
\end{figure}

Along with the local and global order parameters discussed above, two-point
correlation functions afford insight into the organization of the driven
system. The charge-charge and density-density correlation functions are
plotted in Figs.~\ref{gpsi} and \ref{grho}, respectively, for maximum drive.
$G_{\psi ||}$ exhibits the oscillations expected given the stripe pattern.
Interestingly, the decay of $G_{\psi \perp}$, which reflects the persistence
of stripes in the direction perpendicular to the drive, decays in a manner
qualitatively similar to the envelope of the oscillations in $G_{\psi ||}$.
The density-density correlation function along the drive, $G_{\rho ||}$,
exhibits weak oscillations at twice the spatial frequency as the
charge-charge correlation, reflecting a reduction in density at the
interfaces between stripes.

At the lowest density ($\rho=0.7$) shown in the figures, all four
correlation functions decay in a manner best described by a stretched
exponential, $G \sim \exp[-(x/\xi)^\beta]$, with exponents $\beta$ ranging
from 0.36 (for $G_{\rho ||}$) to 0.55 (for $G_{\psi ||}$ and $G_{\psi \perp}$%
. (For correlations along the drive, the \textit{envelope} of the
oscillations decays as a stretched exponential.) For density 0.75,
stretched-exponential decay is again observed, with $\beta \simeq 0.2$,
except in the case of $G_{\psi \perp}$, which is better fit by a power law, $%
G \sim 1/r^\alpha$, with $\alpha = 0.133$. At density 0.8, $G_{\psi ||}$
approaches a nonzero limit of about 0.41 as $x \to \infty$, marking
long-range coherence of the stripe pattern. The other correlation functions
decay as power laws, with exponents $\alpha = 0.080$ (for $G_{\psi \perp}$),
0.62 (for $G_{\rho ||}$) and 0.33 ($G_{\rho \perp}$). Thus, correlations
decay more slowly at higher density. A nonzero limiting value of $G_{\psi
||} $ provides an alternative for detecting long-range order; we defer a
systematic analysis using this criterion to future work.

\subsection{Interface width}

As shown above, the boundary between A- and B-rich phases defines an
interface oriented, on average, perpendicular to the drive, but with
significant fluctuations. Thus the scaling properties of the interface width
with time and system size are of interest. We examine these properties in
the single-stripe regime, in which identification of the interface is
relatively simple. Given a configuration, we define for each $j = 1,..., L$
in the direction perpendicular to the drive, and for $\ell = 1,..., L$ along
the drive, the function,

\begin{equation}
s(\ell,j) \equiv \sum_{i=1}^\ell \sigma_{i,j}.  \label{sellj}
\end{equation}

\noindent Since $s(\ell,j)$ represents the excess of A particles over B
particles in the first $\ell$ sites in row $j$, the values of $\ell$ for
which $s(\ell,j)$ takes its maximum and minimum correspond to the interface
positions in this row. (If the sites at which the maximum or minimum values
occur are not unique, we use the smallest $\ell$ at which the extremum
occurs.) Given the set of sites $\ell_+ (j)$ marking the maximum, we
construct a \textit{random walk representation} of the interface, $h_+(j)$,
by setting $h_+(1) = 0$ and, for $j \geq 2$, taking $h_+(j) = h_+(j-1) +
[\ell_+ (j) - \ell_+ (j-1)]$, where the difference in the square brackets is
calculated under periodic boundaries. A second interface is constructed
using the positions at which $s(\ell,j)$ takes its minimum. The width of
each interface is given by,

\begin{equation}
w = \left\{ \frac{1}{L} \sum_{j=1}^L \left[ h_+(j) - \overline{h} \right]^2
\right\}^{1/2},  \label{width}
\end{equation}

\noindent where $\overline{h} = (1/L) \sum_{j=1}^L h_+(j)$.

\begin{figure}[h!]
\centering
\includegraphics[scale=0.6]{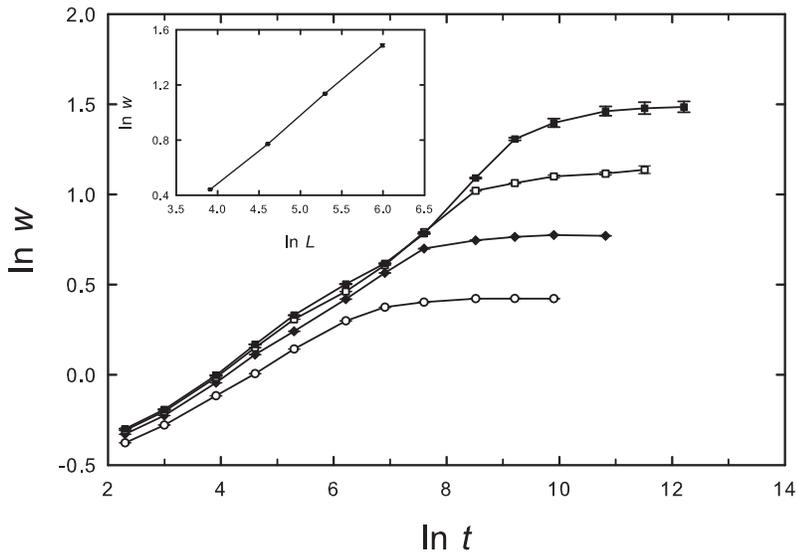}
\caption{Interface width $w$ versus time for $p=1/3$, $a=0$ and system sizes
$L=50$, 100, 200 and 400 (lower to upper). Inset: $w_{sat}$ versus $L$.}
\label{wsc}
\end{figure}

We examine the scaling properties of the interface width as a function of
time and system size, plotting, in Fig.~\ref{wsc} (main graph), the width
versus time for system sizes $L=50$, 100, 200 and 400, and (inset) the
saturation (long-time) width $w_{sat}$ versus system size. These data are
for $p=1/3$, $a=0$, and density $\rho=0.85$ (single-stripe regime). The data
for $w(t)$ do not follow simple power laws, and cannot be collapsed onto a
single curve by rescaling time and width. It is nevertheless possible to
estimate the growth exponent $\beta_w$ (associated with the behavior $w \sim
t^{\beta_w}$) using the most linear portions of the graphs (specifically,
the second through fifth points in the present case), yielding $\beta_w =
0.182(3)$, 0.205(4), 0.223(5) and 0.228(6), for system sizes 50 through 400,
respectively. The expected scaling of the saturation width, $w_{sat} \sim
L^\alpha$ is satisfied to reasonable approximation; for $L \geq 100$ we find
$\alpha = 0.516(6)$. We note that the exponent values are not very different
from those of the Edwards-Wilkinson (EW) class, $\alpha=1/2$ and $\beta_w=1/4$%
; it is conceivable that the observed deviations reflect finite-size
effects. We defer a more precise determination of the interface width,
including analyses of multi-stripe systems, to future work.

\section{Summary and outlook}

We study nonequilibrium stationary properties of a Widom-Rowlinson lattice
gas subject to a drive favoring particle hopping along one axis and
suppressing hopping to the opposite direction. The stationary properties are
surprisingly different from those of the driven lattice gas with attractive
interactions. As in the equilibrium WRLG \cite{wrlg}, there is a phase
transition with segregation between particle species as the density is
increased; the critical density for phase separation increases with drive.
We find that even in the disordered phase, there is a preferred wavelength $%
\lambda$ evident in the charge-charge correlation function along the drive
direction, and that the associated structure factor $S(k,0)$ does not take
the usual Ornstein-Zernike form. With increasing density, local ordering
into A- and B-rich regions occurs, with the appearance of stripes oriented
perpendicular to the drive, but without global coherence. Increasing the
density further, we observe a transition to a stripe pattern with long-range
order. We characterize the system in terms of local and global order
parameters, correlation functions and associated structure factors, and
provide preliminary results on interface roughness and the particle current
provoked by the drive.

Our study leaves many intriguing questions for future study. Among them, we
highlight:

1. Why does phase separation result in stripes of a characteristic width, $%
\lambda$, instead of two regions, as happens in equilibrium? Simulations
suggest that very broad stripes are unstable, but the underlying mechanism
is unclear.

2. Why do the stripes form perpendicular to the drive? Again, we observe an
instability in an interface oriented parallel to the drive - undulations on
such an interface tend to grow - but a quantitative explanation is lacking.

3. Can one develop a hydrodynamic description or time-dependent
Ginzburg-Landau theory capable of reproducing the phenomenology observed in
simulations? Can one derive a quantitative prediction for the current?
We find (Sec.~III) that although simple mean-field analyses
do not predict the phase diagram, they do yield quite good predictions for
the current.

4. What is the nature of the phase transition to long-range order of the
stripe pattern? Can it be connected to (equilibrium) transitions in smectic
liquid crystals?

5. Do the scaling properties of the interfaces fall in the Edwards-Wilkinson
class, or some other known class of growth processes?

6. To what extent does entropy maximization, which is the sole factor
determining static properties at equilibrium, hold in the driven system,
particularly for a weak drive? Related to this, is the zero-drive limit
smooth or singular?

Although these and other questions remain open, our study demonstrates
that the driven Widom-Rowlinson lattice gas exhibits surprising behaviors
not previously observed in driven systems.

\begin{acknowledgments}
This work was supported by CNPq and CAPES,
Brazil.
\end{acknowledgments}


\begin{thebibliography}{99}
\bibitem{notation}
Though we use $\rho \left( \vec{x}\right) $ to denote the local density,
there should be no confusion with $\rho $, the \textit{overall} particle
density. They are related by $\rho L^{2}=\Sigma _{\vec{x}}\rho \left( \vec{x}%
\right) $. Meanwhile, there should be no confusion that \textquotedblleft
charge\textquotedblright\ here is unrelated to the electric charge, but used
as a way to distinguish $A$ from $B$ particles.
\bibitem{beam}
This average power spectrum is precisely the intensity of the scattered
light when a beam is shined on a sample of our fluctuating lattice gas.
\bibitem{finiteL}
Due to both finite $L$ effects and the conservation law (which imposes
non-vanishing values on $G$ at large $r$), extracting asymptotic behaviors
is not straightforward. Thus, we cannot rely on data at distances comparable
to $L/2.$
\end{thebibliography}
\end{document}